\documentclass[aps,prb,reprint,twocolumn,superscriptaddress,notitlepage]{revtex4-1}

\usepackage{amssymb,amsfonts,amsmath}

\usepackage[pdftex]{graphicx}
\usepackage{epstopdf}

\usepackage[utf8]{inputenc}
\usepackage[T1]{fontenc}
\usepackage{color}
\usepackage{setspace}
\DeclareMathOperator{\tr}{Tr}

\newcommand{\ket}[1]{\left | #1 \right\rangle}
\newcommand{\bra}[1]{\left\langle #1 \right |}

\newcommand{\half}{\mbox{$\textstyle \frac{1}{2}$}}

\definecolor{stephen}{rgb}{1,0.1,0.1}

\begin{document}

\title{Non-linear quantum-classical scheme to simulate non-equilibrium strongly correlated fermionic many-body dynamics}



\author{J.~M.\ Kreula}
\altaffiliation{Electronic address: juha.kreula@physics.ox.ac.uk}
\affiliation{Clarendon Laboratory, University of Oxford, Parks Road, Oxford OX1 3PU, United Kingdom}

\author{S.~R.\ Clark}
\affiliation{Department of Physics, University of Bath, Claverton Down, Bath BA2 7AY,  United Kingdom}
\affiliation{Max Planck Institute for the Structure and Dynamics of Matter, Hamburg, Germany}

\author{D.\ Jaksch}
\affiliation{Clarendon Laboratory, University of Oxford, Parks Road, Oxford OX1 3PU, United Kingdom}
\affiliation{Centre for Quantum Technologies, National University of Singapore, 3 Science Drive 2, Singapore 117543}

%\date{\today}
\begin{abstract}
We propose a non-linear, hybrid quantum-classical scheme for simulating non-equilibrium dynamics of strongly correlated fermions described by the Hubbard model in a Bethe lattice in the thermodynamic limit. Our scheme implements non-equilibrium dynamical mean field theory (DMFT) and uses a digital quantum simulator to solve a quantum impurity problem whose parameters are iterated to self-consistency via a classically computed feedback loop where quantum gate errors can be partly accounted for. We analyse the performance of the scheme in an example case.
\end{abstract}

\maketitle
\onecolumngrid
%Introduction
Next generation scalable quantum devices\cite{nqit,barends2015digital} promise a step change in our ability to do computations. Direct quantum simulation\cite{feynman1982simulating,buluta2009quantum,johnson2014quantum} using highly controllable quantum systems~\cite{blatt2012quantum,bloch2012quantum,houck2012chip} has already led to numerous insights into many-body quantum physics, despite limitations in the size of the simulated system.

Recently, quantum computer simulations of strongly correlated fermion models have been proposed~\cite{PhysRevA.92.062318,PhysRevA.93.032303}. We suggest a hybrid quantum-classical scheme to simulate non-equilibrium dynamics of the Hubbard model in a Bethe lattice directly in the thermodynamic limit. Our scheme implements the non-equilibrium extension of the well-established dynamical mean-field theory (DMFT) method (for extensive reviews of DMFT, see, e.g., Refs.~\cite{georges1996dynamical,RevModPhys.86.779}). Instead of the traditional all-classical method, the proposed scheme uses a digital quantum simulator to efficiently solve the DMFT impurity problem, the parameters of which are iterated to self-consistency via a classically computed feedback loop. This setup promises an exponential speed-up over the best currently-known Hamiltonian-based classical algorithms.  We show how quantum gate errors can be partly accounted for in the feedback loop, improving simulation results. The scheme also avoids the sign problem in classical quantum Monte Carlo methods and works for all interaction strengths, unlike classical methods based on perturbation theory. Presently, non-equilibrium DMFT is one of the most promising methods to study time-dependent phenomena in high-dimensional correlated lattice models, and could thus be of interest for current efforts to develop scalable quantum technologies\cite{nqit,benhelm2008towards,lanyon2011universal,blatt2012quantum}. Examples of applications of non-equilibrium DMFT include the dielectric breakdown of Mott insulators~\cite{eckstein2010dielectric}, damping of Bloch oscillations~\cite{eckstein2011damping}, and thermalization after parameter quenches~\cite{PhysRevLett.103.056403,PhysRevB.81.115131}.

%Motivation
Further to this, driven strongly correlated quantum materials are now being extensively investigated experimentally. A large motivation for this is the possibility of manipulating correlated phases of matter with strong pulses of light, such as photodoping of Mott insulators~\cite{wall2011quantum}  or inducing superconductivity~\cite{fausti2011light}. The underlying physical mechanisms are, however, still poorly understood. Even the dynamical behaviour of conceptually simple and commonly used quantum lattice models is yet not fully grasped. Solving these model systems could elucidate physical phenomena underlying currently unexplained experimental results. A standard example of this kind of idealised model for non-equilibrium problems is the time-dependent Hubbard Hamiltonian
 \begin{align}\label{eq:hubbard}
\hat{H}(t)=-v(t)\sum_{\langle i, j \rangle \sigma} \left( \hat{c}^{\dagger}_{i,\sigma} \hat{c}_{j,\sigma} + \mathrm{H.c.} \right)+ U(t) \sum_i \left( \hat{n}_{i,\downarrow} - \frac{1}{2}\right)\left( \hat{n}_{i,\uparrow} - \frac{1}{2}\right) .
\end{align}
In this model, electrons with spin projections $\sigma=\downarrow, \uparrow$ move only between adjacent lattice sites $i$ and $j$ with time-dependent `hopping' energy $v(t)$, where $t$ denotes time. This process is described in the first sum, which is over all nearest-neighbour sites, with fermionic creation and annihilation operators $\hat{c}^{\dagger}_{i,\sigma}$ and $\hat{c}_{j,\sigma}$, respectively. The electrons interact with Coulomb repulsion $U(t)$ only if they occupy the same lattice site $i$, given in the latter term by the product of the number operators $\hat{n}_{i,\downarrow}=\hat{c}^{\dagger}_{i,\downarrow}\hat{c}_{i,\downarrow}$  and $\hat{n}_{i,\uparrow}=\hat{c}^{\dagger}_{i,\uparrow}\hat{c}_{i,\uparrow}$.

\begin{figure*}[ht!]
\centerline{\includegraphics[scale=0.4]{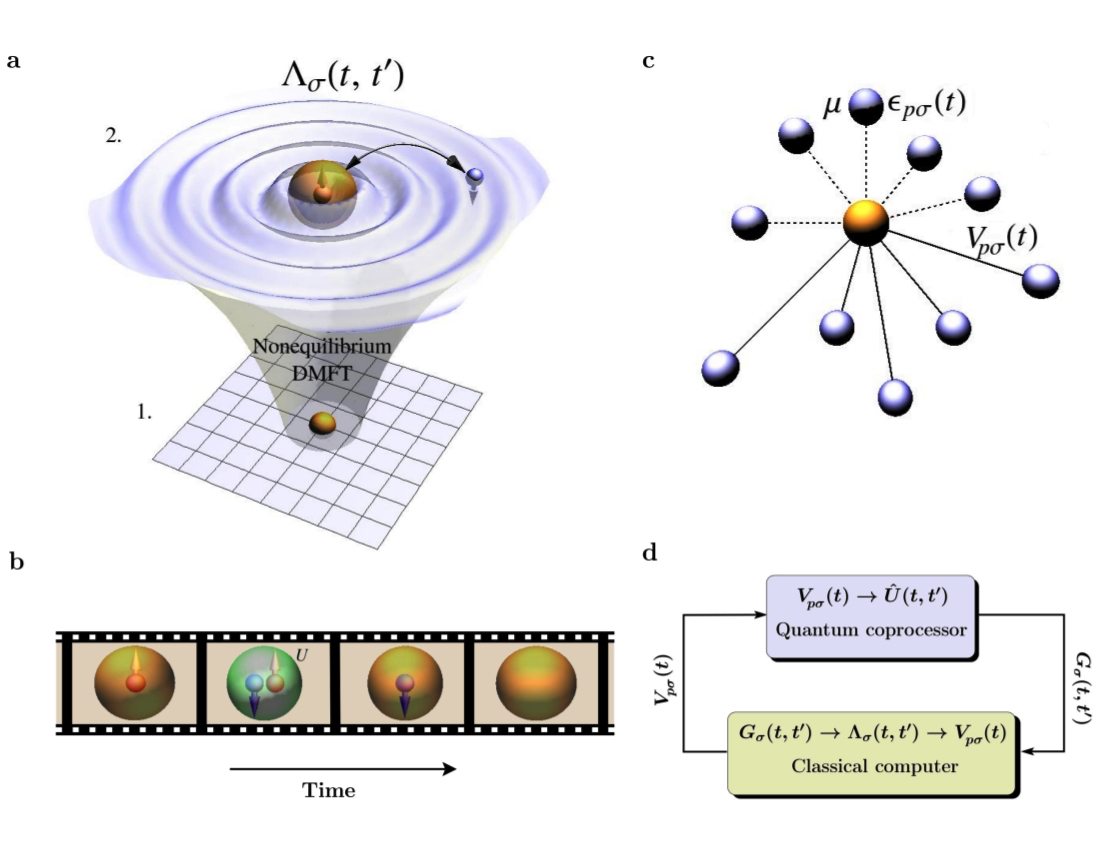}}
\caption{\textbf{a.} In non-equilibrium DMFT a fermionic quantum lattice model is replaced by a single impurity site exchanging particles via a self-consistently determined time and spin dependent mean field $\Lambda_\sigma(t,t')$. \textbf{b.} This exchange of particles yields dynamical fluctuations of the impurity site occupation as a function of time shown here as $|\uparrow \rangle \rightarrow  | \downarrow  \uparrow \rangle \rightarrow |\downarrow \rangle \rightarrow |{\rm vac}\rangle$. The onsite interaction $U$ energetically penalises the doubly occupied state $| \downarrow  \uparrow \rangle$. \textbf{c.} The impurity-mean field interaction is mapped onto a SIAM with unitary evolution $\hat U(t,t')$. The energies of the non-interacting bath sites $p$ are chosen $ \epsilon_{p\sigma}(t)=0$ for $t > 0$ and their chemical potential is set $\mu=0$ in this work \cite{gramsch2013hamiltonian}. The impurity site exchanges fermions with time-dependent hybridization energies $V_{p\sigma}(t)$. \textbf{d.} Quantum-classical hybrid simulation scheme: the SIAM dynamics for a given set of parameters $V_{p\sigma}(t)$ is implemented on a quantum coprocessor and yields the impurity Green's function $G_{\sigma}(t,t')$. The classical non-linear feedback loop takes $G_{\sigma}(t,t')$ and calculates the mean field $\Lambda_\sigma(t,t')$ from which a new set of $V_{p\sigma}(t)$ can be extracted. These parameters are then fed back into the quantum coprocessor and the loop is repeated until self-consistency is achieved.}
\label{fig:dmft}
\end{figure*}

This and similar models are extremely challenging to study numerically due to the exponential growth of the Hilbert space with system size. One thus often resorts to mean field approximations which typically consider only a single lattice site and replace interactions with its neighbourhood by a mean field $\Lambda$. This turns a linear quantum problem in an exponentially large Hilbert space into a much smaller but non-linear problem where $\Lambda$ needs to be determined self-consistently. Such mean field approximations become increasingly accurate with the number of nearest neighbours. A classic example of this approach is the Weiss theory of ferromagnetism \cite{cardy1996scaling}. For mean field theory to be applicable to strongly correlated Fermi systems in thermal equilibrium, the mean field $\Lambda_\sigma(t)$ has to be dynamical to account for correlations between interactions with the environment that are separated by $t$ in time, as schematically shown in Figs.~\ref{fig:dmft}a,b. 

This highly successful approach is called DMFT~\cite{georges1996dynamical}. DMFT can be extended to non-equilibrium systems~\cite{RevModPhys.86.779} by letting $\Lambda_\sigma(t,t')$, which is often called hybridization function, depend on two interaction times $t$ and $t'$ explicitly. Note that non-local spatial fluctuations can be included in DMFT by going beyond the single-site approximation and considering a cluster of isolated sites~\cite{RevModPhys.77.1027,PhysRevB.90.075117}, but this is beyond the scope of this work.

%Classical DMFT algorithm
In general, it is a complex task to determine $\Lambda_{\sigma}(t,t')$ and the related local single-particle Green's function {$G_{\sigma}(t,t')=-i\langle \mathcal{T} \hat{c}_{\sigma}(t) \hat{c}^{\dagger}_{\sigma}(t')  \rangle$ (where $\mathcal{T} $ is the time-ordering operator)}, describing the response of the many-body system after a localized removal and addition of a particle at times $t$ and $t'$. Commonly used numerical methods for solving the non-equilibrium DMFT problem include continuous-time quantum Monte Carlo, which suffers from a severe dynamical sign problem, and perturbation theory which can only address the weak and strong coupling regimes~\cite{RevModPhys.86.779}. 

In infinite dimensions, the system can also be explicitly mapped onto a single impurity Anderson model (SIAM)~\cite{gramsch2013hamiltonian} 
\begin{align}\label{eq:SIAM}
\hat{H}_{\mathrm{SIAM}}(t)=\hat{H}_{\mathrm{loc}}(t)+\hat{H}_{\mathrm{bath}}(t)+\hat{H}_{\mathrm{hyb}}(t),
\end{align}
\begin{align}
\hat{H}_{\mathrm{loc}}(t)= U(t)\, \left(\hat{n}_{\uparrow}-\frac{1}{2}\right) \left(\hat{n}_{\downarrow}-\frac{1}{2}\right)-\mu \sum_\sigma \hat{n}_{\sigma},
\end{align}
\begin{align}
\hat{H}_{\mathrm{hyb}}(t)=\sum_{p} \left( V_{p\sigma}(t)\hat{c}_{\sigma}^{\dagger}\hat{c}_{p\sigma} + \mathrm{H.c.} \right),
\end{align}
\begin{align}
\hat{H}_{\mathrm{bath}}(t)=\sum_{p,\sigma} \left[ \epsilon_{p\sigma}(t)-\mu \right]\hat{c}_{p\sigma}^{\dagger}\hat{c}_{p\sigma}.
\end{align}
where the selected lattice site is represented by an impurity, with the creation (annihilation) operator $\hat{c}_{\sigma}^{\dagger}$ ($\hat{c}_{\sigma}$) and number operator $\hat{n}_\sigma=\hat{c}_{\sigma}^{\dagger}\hat{c}_{\sigma}$, whose interaction with $\Lambda_{\sigma}(t,t')$ is mimicked by a collection of $N$ non-interacting bath sites with on-site energies $\epsilon_{p\sigma}(t)$, as shown in Fig.~\ref{fig:dmft}c. The time-dependent hybridization energy $V_{p\sigma}(t)$ describes the amplitude for exchange of fermions between the impurity site and bath site $p$. These must be determined self-consistently: for given $V_{p\sigma}(t)$ the quantum dynamics of the SIAM is solved and its Green's function and corresponding {hybridization function} $\Lambda_{\sigma}(t,t')$ are determined. From $\Lambda_{\sigma}(t,t')$ a new set of $V_{p\sigma}(t)$ is worked out which is then fed back into the SIAM. These steps are repeated until convergence is achieved \cite{gramsch2013hamiltonian}. The dynamics of the SIAM is usually worked out {with} exact diagonalization (ED)~\cite{gramsch2013hamiltonian} for small systems or with tensor network theory (TNT) methods\cite{PhysRevB.90.235131}. However, the dynamical generation of entanglement in these problems has severely hampered the efficiency of TNT methods \cite{cirac2012goals,PhysRevB.90.235131}. Furthermore, the required number of bath sites increases with the maximum simulation time $t_{\rm max}$. This makes solving the SIAM the exponentially difficult bottleneck~\cite{gramsch2013hamiltonian,PhysRevB.90.235131,PhysRevB.91.045136} in purely classical DMFT solvers.

%Brief summary of work
Here, we propose and analyze a hybrid quantum-classical computing scheme for DMFT to efficiently solve the Hubbard model in a Bethe lattice. The Bethe lattice is chosen for the simplicity of its self-consistency condition. It is conceptually straightforward to extend the scheme to other types of lattices. A small digital quantum coprocessor solves the SIAM evolution with the resulting $G_{\sigma}(t,t')$ being processed by a classical computer to complete the non-linear feedback loop as shown in Fig.~\ref{fig:dmft}d. We consider a trapped ion coprocessor for concreteness, although any other platform for quantum computing could implement the coprocessor as well. Even for imperfectly implemented quantum gates with realistic errors of $1\%$ we find accurate solutions to a simple model problem in small systems. In addition, our numerical evidence suggests that gate errors mainly lead to a smearing of the bath energies, which can be accounted for in the classical feedback loop to improve the solution.

\begin{figure}[ht!]
\includegraphics[scale=0.4]{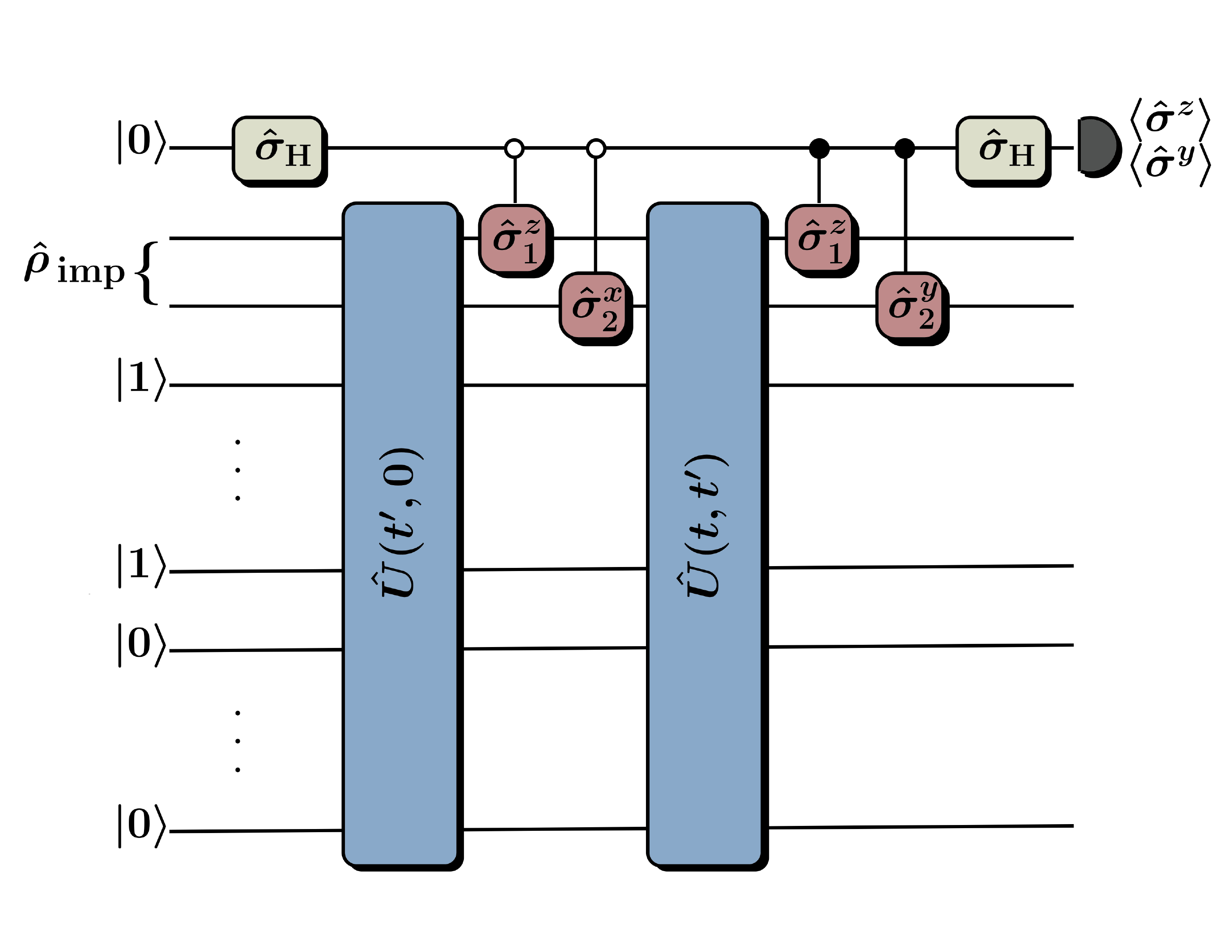}
\caption{Coprocessor quantum network for measuring a contribution to $G_\sigma(t,t')$ in the SIAM dynamics. This example network is given for the paramagnetic phase starting from the atomic limit, as considered in the main text and in Ref.\cite{gramsch2013hamiltonian}. A probe qubit (top line) is prepared in a symmetric superposition $(|0\rangle + |1\rangle)/\sqrt{2}$ of computational basis states $|0\rangle$ and $|1\rangle$ by a Hadamard gate $\hat \sigma_H$. Here, $\hat{\rho}_{\rm imp}=\frac{1}{2}\left( | 0, 1 \rangle \langle 0,1 | + | 1,0 \rangle \langle  1,0  |\right)$, and the initial states of the bath sites  (lines below the impurity) are set to either $|0\rangle$ or $|1\rangle$ using Jordan--Wigner transformed operators, following the standard scheme in Ref.\cite{gramsch2013hamiltonian}. After evolving the SIAM to time $t'$ according to $\hat{U}(t',0)$ the probe qubit interacts with the impurity via controlled Pauli gates. A second set of controlled Pauli gates is applied after evolving the impurity to time $t$. The precise choice of Pauli gates selects different contributions to the Green's function. After another Hadamard gate this contribution is encoded in the expectation values $\hat{\sigma}^z$ and $\hat{\sigma}^y$ of the probe qubit, as discussed in Methods.}
\label{fig:co}
\end{figure}

%Hybrid Coprocessor scheme
Figure~\ref{fig:co} shows an example coprocessor quantum network for computing a contribution to the Green's function (see Methods for details). The real and imaginary contributions to the impurity Green's function are encoded as $\langle \sigma^z\rangle$ and $\langle \sigma^y\rangle$ of a probe qubit by interacting it with the impurity state at times $t'$ and $t$ via controlled quantum gates\cite{dorner2013extracting}. We decompose the unitary dynamics $\hat{U}(t,t')$ of the SIAM into a network of quantum gates \cite{casanova2012quantum,muller2011simulating} by discretising time as $t_n = n\Delta t$, where $\Delta t$ is a small time-step. We then breakup the evolution from $t=0$ to $t=t_n$ into a product of Trotter steps $\hat{U}(t_n,0) = \prod_{l=0}^{n-1} \hat{U}(l\rightarrow l+1)$. The Trotter steps can readily be implemented by single qubit rotations and multi-qubit entangling M{\o}lmer--S{\o}rensen (MS) gates~\cite{molmer1999multiparticle,muller2011simulating} that have recently been realized in ion traps with high fidelity~\cite{benhelm2008towards,lanyon2011universal}. The total number of MS gates per Trotter step scales only linearly with the number of bath sites.

\begin{figure*}[ht!]
\centerline{
\includegraphics[scale=0.9]{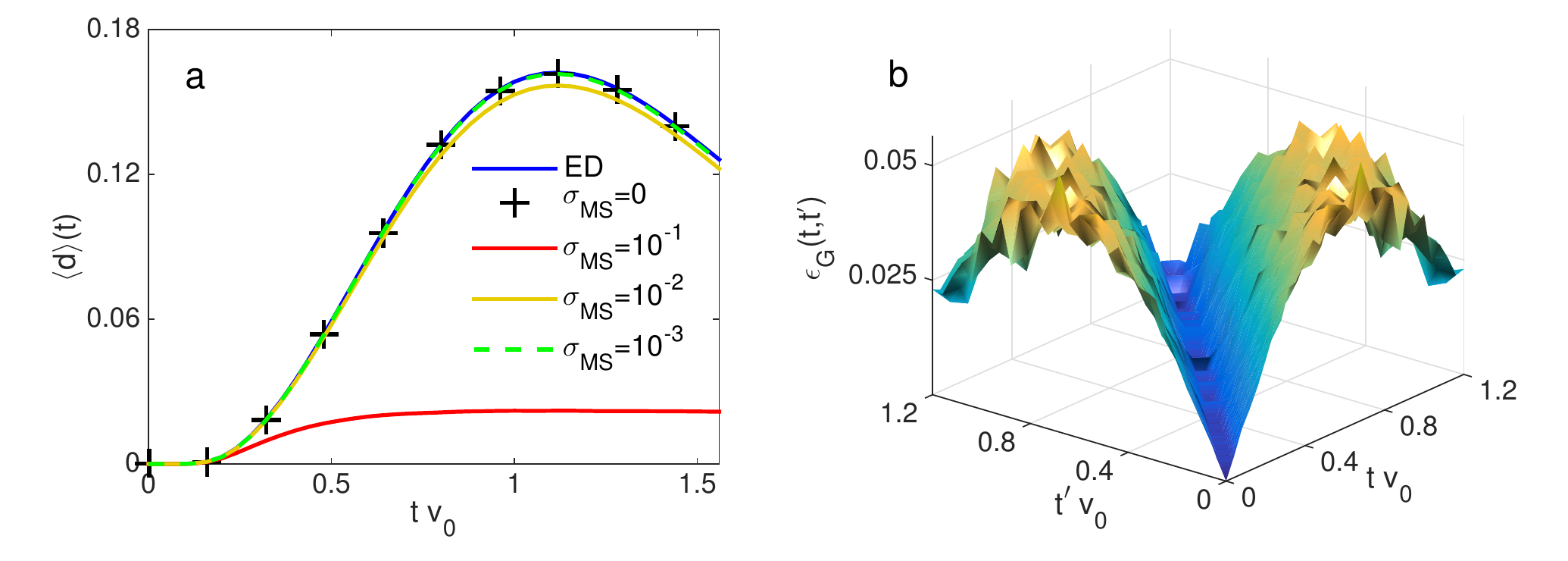}
}
\caption{Hybrid non-equilibrium DMFT simulation results when dynamically increasing the Hubbard tunneling matrix element $v(t)$ from $0$ to $v_0$ as described in the main text. We choose $U=2 v_0$, Trotter steps $\Delta t = 0.04/v_0$ and couple the impurity site to $N=2$ bath sites. \textbf{a.} Impurity double occupation $\langle \hat{d} \rangle(t)$ as a function of time $t$: numerically exact solution (blue solid curve), solution with Trotter errors $(+)$, solutions including gate errors of $\sigma_{\rm MS}=0.1\%$ (green dashed curve), $\sigma_{\rm MS}=1\%$ (yellow solid curve), and $\sigma_{\rm MS}=10\%$ (red solid curve). \textbf{b.} Absolute value of the difference $\epsilon_G(t,t')$ between the imaginary parts of the lesser Green's function without gate errors and with gate errors of $\sigma_{\rm MS}=1\%$. Results of calculations with gate errors are obtained by averaging over $128$ realizations {of the setup}.}
\label{fig:res1}
\end{figure*}

%Example calculation
We analyze the performance of our simulation scheme by considering a simple example system~\cite{gramsch2013hamiltonian}. We study the infinite-dimensional time-dependent Hubbard model~\eqref{eq:hubbard} with constant onsite interaction $U$ and tunneling matrix element $v(t)$. The simulation starts in the half-filled paramagnetic atomic limit with tunneling $v(t=0)=0$, which is then dynamically ramped up to its final value $v_0$ after quench time $1/4v_0$ and is kept at $v_0$ until the final simulation time $t_{\rm max}$ is reached\cite{gramsch2013hamiltonian} (setting $\hbar=1$). Such a sudden quench is representative of experimental ultracold atom dynamics~\cite{esslinger2010,langen2015ultracold} and also ultrafast dynamics probed in condensed matter systems~\cite{wall2011quantum}. The initial state of the system has a singly occupied impurity site in the completely mixed state of spin $\uparrow$ and spin $\downarrow$, and one half of the bath sites are doubly occupied and the other half empty (for explicit details, see~Ref.\cite{gramsch2013hamiltonian}). In practice, we prepare the system in two pure fermion occupational number states, where one has the impurity in state $|\uparrow\rangle$ and the other in state $|\downarrow\rangle$, along with the bath states~\cite{gramsch2013hamiltonian}. The results are then averaged over these two pure states. These initial number states are mapped onto product states of qubits via the Jordan--Wigner transformation (see Methods). The initial qubit configuration is that shown in Fig.~\ref{fig:co}, where $\hat{\rho}_{\rm imp}=\frac{1}{2}\left( | 0, 1 \rangle \langle 0,1 | + | 1,0 \rangle \langle  1,0  |\right)$. We emulate the operation of the quantum coprocessor by classically evaluating the quantum networks, and the classical exponential scaling limits our simulations to small systems. The self-consistency condition for the Bethe lattice calculated in the classical feedback loop is $\Lambda_{\sigma}(t,t')=v(t)G_{\sigma}(t,t')v(t')$, from which we obtain the SIAM coupling to bath $p$ efficiently via a Cholesky decomposition $\Lambda_{\sigma}(t,t') = \sum_p V_{p\sigma}(t) V^*_{p\sigma}(t')$, where $^*$ denotes complex conjugation (see Supplementary Material for details). The impurity site double occupancy $\langle \hat{d} \rangle(t) = \langle \hat{n}_{\downarrow} \hat{n}_{\uparrow} \rangle(t)$ obtained from the self-consistent hybrid simulation is compared to the exact result in Fig.~\ref{fig:res1}a and shows that Trotter errors do not noticeably affect our results.

%Imperfect gates
Next we assume imperfect gates characterized by phase errors that are described by normally distributed random variables with zero mean~\cite{PhysRevA.69.042314}. We choose their standard deviations consistent with current experimental capabilities\cite{harty2014high,benhelm2008towards,nqit} setting the single qubit error to $\sigma=10^{-6}$ and allowing MS gate errors $\sigma_{\rm MS}$ to vary between $0.1\%$ and $10\%$. We obtain accurate results for the dynamics of the double occupancy even in the presence of gate errors. As shown in Fig.~\ref{fig:res1}a the double occupation differs from the exact result by only $\approx 3\%$ for $\sigma_{\rm MS}=1\%$. For a smaller gate error of $\sigma_{\rm MS}=0.1\%$ the difference is insignificant up to $t=1.5/v_0$. In Fig.~\ref{fig:res1}b we plot the error in the imaginary part of the lesser Green's function $G^<_{\sigma}(t,t')$ induced by imperfect gates. The diagonal values $G^<_\sigma(t,t)$, which determine time-local single-particle observables, are almost unaffected even for large MS gate errors. Gate errors in general make the Green's function decay faster with $t-t'$ than in the ideal case and will thus affect unequal time correlation functions.

\begin{figure*}[ht!]%
\centerline{
\includegraphics[scale=0.9]{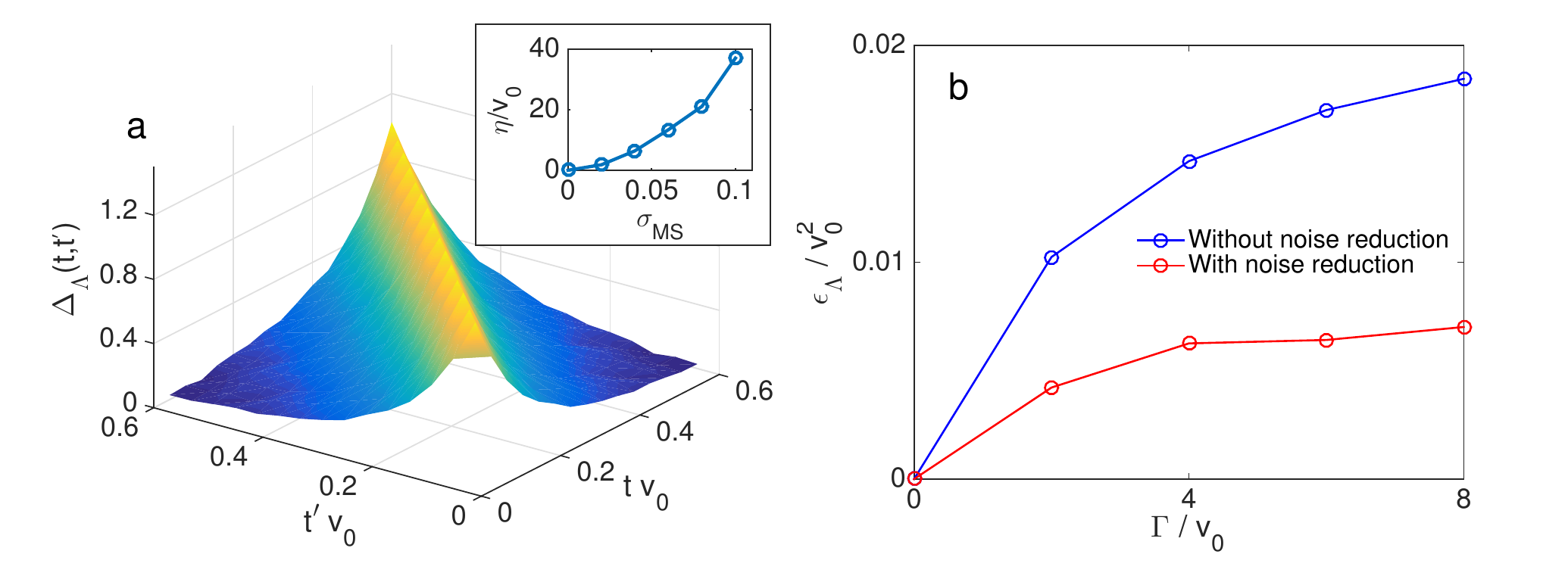}}
\caption{\textbf{a.} Deviation of the mean field $\Delta_\Lambda(t,t') = {\rm Im}\Lambda^{<}_{\eta}(t,t')/{\rm Im}\Lambda^{<}_{0}(t,t') \approx \exp(-\eta |t-t'|)$, where $\Lambda^{<}_{(0,\eta)}(t,t')$ is the lesser component of the mean field, in the absence (presence) of gate errors (of  $\sigma_{\rm MS}=6\%$) for constant hybridizations and $U=2 v_0$, $N=2$ and averaged over $128$ realizations. The inset shows the exponential decay rate $\eta$ against two qubit error $\sigma_{\rm MS}$. \textbf{b.} Average error in the self-consistent mean field $\epsilon_\Lambda=|\Lambda^<_{\rm noisy}(t,t')-\Lambda^<_{\rm exact}(t,t')|$ for the non-interacting system with $N=10$ noisy bath sites.}
\label{fig:res2}
\end{figure*}

%Effect of gate errors
We further investigate the effect of imperfect gates by considering the impurity site coupled to two bath sites via constant $V_{p\sigma}(t)$. We find that the imaginary part of the mean field differs from the exact solution by a factor of approximately $\exp(-\eta |t'-t|)$ as shown in Fig.~\ref{fig:res2}a. The decay rate $\eta$ increases with $\sigma_{\rm MS}$ as displayed in the inset of Fig.~\ref{fig:res2}a. This numerical evidence suggests that gate errors have the same effect as smearing out the bath energies $\epsilon_{p\sigma}(t)$ to a similar width $\eta$. The impurity model including errors would then be equivalent to the bath sites possessing a finite coherence time $1/\eta$. Since the number of gates is $\propto N$ we expect $\eta$ to only depend weakly on $N$.

%Modified relation between \Lambda and V
A bath site with coherence time $1/\eta$ can be modelled by allowing an ideal bath to incoherently exchange particles with a reservoir at an `error' rate $\Gamma = \eta$. This exchange of particles modifies the bath's Green's function from its ideal value of $g_{p\sigma}(t,t')=1$ and correspondingly modifies the relation between impurity bath couplings and mean field to~\cite{gramsch2013hamiltonian} $\Lambda_{\sigma}(t,t') = \sum_{p} V_{p\sigma}(t) g_{p\sigma}(t,t') V^*_{p\sigma}(t')$. This relation does not necessarily allow for an exact solution for $V_{p\sigma}(t)$ even for large $N$. The effect of noise therefore limits the mean fields $\Lambda_{\sigma}(t,t')$ that the bath sites can model.

%Noise reduction in simplified model
We investigate if the noise induced by gate errors can be partly compensated by implementing self-consistency via this modified relation. For the non-interacting impurity with bath sites coupled to a particle reservoir we solve numerically for the bath Green's functions $g_{p\sigma}(t,t')$, exploiting the super-fermion formalism~\cite{superfermion} (see Supplementary Material). We minimize $\left \| \sum_{p} V_{p\sigma}(t) g_{p\sigma}(t,t')V^*_{p\sigma}(t')- \Lambda_{\sigma}(t,t') \right \|$ using the Frobenius norm over the $V_{p\sigma}(t)$ to obtain the hybridizations in the noisy system. This modification of the {\em classical} feedback loop significantly reduces the effect of gate errors as demonstrated in Fig.~\ref{fig:res2}b, showing the reduction in average absolute error in the mean field $\Lambda_{\sigma}(t,t')$. In the hybrid simulation scheme a slight modification of the quantum network shown in Fig.~\ref{fig:co} allows the probe qubit to measure the bath Green's functions, thus providing the information required for this noise-reduction scheme to be implemented.

%Conclusions
Finally, we emphasize that our scheme works directly in the thermodynamic limit and, since it does not require a small expansion term, gives accurate results for all values of $U$, in particular for the challenging situation of intermediate interactions like the example $U=2 v_0$ considered here. The number of available qubits only limits the number of bath sites that can be included in the simulation and hence the maximally reachable simulation time $t_{\rm max}$. Purely classical simulations are currently limited to approximately $25$ bath sites\cite{PhysRevB.90.235131} and, because of fast growing SIAM entanglement\cite{gramsch2013hamiltonian,PhysRevB.90.235131}, scale exponentially with $t_{\rm max}$ despite efficiently implementing the feedback loop. Therefore, a quantum coprocessor with only about $50$ qubits\cite{nqit} coupled to a classical feedback loop would be able to improve upon current purely classical algorithms. Our hybrid simulation scheme thus provides an interesting scientific application of next generation, possibly imperfect, quantum devices.

The authors would like to thank Simon Benjamin as well as Ian Walmsey and his group members for useful discussions. The research leading to these results has received funding from the EPSRC National Quantum Technology Hub in Networked Quantum Information Processing (NQIT). J.M.K.\ acknowledges financial support from Christ Church, Oxford and the Osk.\ Huttunen Foundation. D.J.\ acknowledges financial support from the European Research Council under the European Union's Seventh Framework Programme (FP7/2007-2013)/ERC Grant Agreement no.\ 319286 Q-MAC and the EU Collaborative project QuProCS (Grant Agreement 641277). The data presented in this work is contained in the source files of the arXiv submission arXiv:1510.05703.

While preparing this manuscript, we became aware of related work by B.\ Bauer et al~\cite{bauer}.

The authors declare no competing financial interests.

Author contributions: J.M.K. and S.R.C. carried out the numerical calculations. J.M.K. decomposed the SIAM into quantum networks and carried out the analytical calculations. D.J., S.R.C. and J.M.K. wrote the manuscript. D.J. conceived and coordinated the project.

\section*{Methods}

\subsection*{Implementing the single-impurity Anderson model with the digital quantum simulator}
To implement the SIAM in Eq.~\eqref{eq:SIAM} in the main text with the digital quantum simulator, we first map the creation and annihilation operators in $\hat{H}_{\mathrm{SIAM}}(t)$ onto spin operators that act on the qubits in the coprocessor. This is achieved via the Jordan--Wigner transformation $\hat{c}^{\dagger}_{p\downarrow} = \bigotimes_{j=1}^{2p-2}\hat{\sigma}_{j}^z \otimes \hat{\sigma}^{-}_{2p-1}$, $\hat{c}^{\dagger}_{p\uparrow} = \bigotimes_{j=1}^{2p-1}\hat{\sigma}_{j}^z \otimes \hat{\sigma}^{-}_{2p}$, and $\hat{c}_{p\sigma}=(\hat{c}^{\dagger}_{p\sigma})^{\dagger}$ (we take $p=1$ to be the impurity). Here,  $\hat{\sigma}^{\pm}=\frac{1}{2}(\hat{\sigma}^x \pm i \hat{\sigma}^y)$, and $\hat{\sigma}^x$, $\hat{\sigma}^y$, and $\hat{\sigma}^z$ are the Pauli spin operators. The transformation maps $N$ fermionic sites onto a string of $2N$ qubits such that two adjacent qubits represent one lattice site. The correspondences between the qubit states and fermionic states are $| 0, 0 \rangle = | {\rm vac} \rangle$, $| 1, 0 \rangle = |\downarrow \rangle$, $|0,1\rangle = |\uparrow \rangle$, and $| 1,1 \rangle = |\downarrow \uparrow \rangle$.

To obtain the necessary quantum gates to approximate the unitary evolution operator we use a Trotter decomposition on the propagator $\hat{U}(n\rightarrow n+1)$ between each time $t_n$ and $t_{n+1}$ as $\hat{U}(n\rightarrow n+1)=e^{-i\Delta t\hat{H}_{\mathrm{SIAM}}(t_n)} \approx \prod_{j} e^{-i\Delta t \hat{H}_j(t_n)}$, where $\hat{H}_{\mathrm{SIAM}}(t_n) =\sum_j \hat{H}_j(t_n)$. Each term $e^{-i\Delta t \hat{H}_j(t_n)}$ can be readily implemented using spin rotations $\hat{U}_{\mathrm{rot}}(\varphi)$ where $\varphi$ is the angle of rotation, and multi-qubit M{\o}lmer--S{\o}rensen (MS) gates~\cite{molmer1999multiparticle,muller2011simulating}, characterized by two phases $\theta$ and $\phi$ as $
\hat{U}_{\mathrm{MS}}^{l,m}\left( \theta, \phi \right)=\exp \left[ -i\frac{\theta}{4}\left(\cos\phi \, \hat{S}_x + \sin\phi \hat{S}_y \right)^2 \right]$,
with $\hat{S}_{x,y}=\sum_{j=l}^m \hat{\sigma}_j^{x,y}$ (see Supplementary Material). Here, the MS gate acts on qubits $l$, $l+1$, \dots, $m$, and the phase $\theta$ controls the amount of entanglement, while varying $\phi$ allows a shift between a $\hat{\sigma}^x$ or a $\hat{\sigma}^y$ type gate.

\subsection*{Measuring the impurity Green's function with single-qubit interferometry}
Using the Jordan--Wigner transformation, the lesser and greater impurity Green's functions for each spin $\sigma$ can be written as a sum of four expectation values of products of Pauli operators and evolution operators (see Supplementary Material). We use a single-qubit interferometry scheme\cite{dorner2013extracting} to measure each of the expectation values $F(t,t')$ that constitute the Green's function. We introduce a probe qubit which is coupled to the string of $2N$ system qubits. We assume that the probe qubit is prepared in the pure state $|0\rangle$, yielding the total system-probe density operator $\hat{\rho}_{\mathrm{tot}}=\hat{\rho}_{\mathrm{sys}}\otimes |0\rangle \langle 0|$. The combined system is then run through a Ramsey interferometer sequence, in which first a $\pi/2$ pulse (or Hadamard gate $\hat{\sigma}_H$) is applied to the probe qubit, the state of which will transform into the superposition $\left(|0\rangle + |1\rangle \right)/\sqrt{2}$. The two states in the superposition provide the necessary interference paths. Following the $\pi/2$ pulse, we apply the unitary evolution on the system of interest up to a certain time $t'$. The Pauli operators are then applied on the system as controlled quantum gates with either $|0\rangle$ or $|1\rangle$ as the control state. This is followed by evolution up to the final time $t'$, another controlled application of Pauli gates, and finally another $\pi/2$ pulse is applied on the probe qubit, bringing the interference paths together. The output state of the probe qubit at the end of the Ramsey sequence is given by
\begin{align}
\hat{\rho}_{\mathrm{probe}}=&\tr_{\mathrm{sys}}\left[\hat{\sigma}_H \hat{T} \hat{\sigma}_H \hat{\rho}_{\mathrm{tot}} \hat{\sigma}_H \hat{T}^{\dagger} \hat{\sigma}_H \right]\nonumber \\
=&\frac{1+\mathrm{Re}[F(t,t')]}{2}|0\rangle \langle 0 | - i\frac{ \mathrm{Im}[F(t,t')]}{2}|0 \rangle \langle 1 |  + i\frac{ \mathrm{Im}[F(t,t')]}{2}|1 \rangle \langle 0 | + \frac{1-\mathrm{Re}[F(t,t')]}{2}|1 \rangle \langle 1 |,
\end{align}
where
%\begin{align}
$F(t,t')=\tr_{\mathrm{sys}}\left[ \hat{T}_{1}^{\dagger}(t) \hat{T}_{0}(t,t') \hat{\rho}_{\mathrm{sys}} \right].$
%\end{align}
Here, the unitary operators $\hat{T}_{0}(t,t')=\langle 0 | \hat{T} | 0 \rangle=\hat{U}(t,t')  \hat{\sigma} \hat{U}(t',0)$ and $\hat{T}_{1}(t)=\langle 1 | \hat{T} | 1 \rangle = \hat{{\sigma}}' \hat{U}(t,0)$, in which $\hat{\sigma}$ and $\hat{{\sigma}}'$ are Pauli operators or tensor products of Pauli operators (see Supplementary Material), act only on the system and not on the probe qubit. Note that we can write
%\begin{align}
$\hat{\rho}_{\mathrm{probe}}=\frac{1}{2}\left(\hat{I} + \mathrm{Re}[F(t,t')]\hat{\sigma}_z +\mathrm{Im}[F(t,t')] \hat{\sigma}_y \right),$
%\end{align}
so that we have
%\begin{align}
$\tr_{\mathrm{probe}}\left[\hat{\rho}_{\mathrm{probe}} \hat{\sigma}_z \right]=\mathrm{Re}[F(t,t')],$
%\end{align}
and
%\begin{align}
$\tr_{\mathrm{probe}}\left[\hat{\rho}_{\mathrm{probe}} \hat{\sigma}_y \right]=\mathrm{Im}[F(t,t')]$.
%\end{align}
Therefore repeated measurements (which can be done in parallel) of the $\hat{\sigma}_z$ and $\hat{\sigma}_y$ components of the probe qubit for all times $t'$ and $t$ yields a contribution to the impurity Green's function $G_{\sigma}(t,t')$. For a spin-symmetric system, on the order of 80,000 measurements per time step are required. See Supplementary Material for details.

%\bibliography{scirepbib}

%\end{thebibliography}

\clearpage
\onecolumngrid

\section*{\large Supplementary material to: ``Non-linear quantum-classical scheme to simulate non-equilibrium strongly correlated fermionic many-body dynamics''}

We provide background details on non-equilibrium dynamical mean-field theory and the setup that we study in the main text, and present the formulae needed for implementing the single-impurity Anderson model with qubits and on measuring the single-particle non-equilibrium Green function. We further elucidate the non-interacting system where we studied reducing the effects of a noisy bath.
\vspace{0.5cm}
\setcounter{equation}{0}
%\end{abstract}
\twocolumngrid

\section{Non-equilibrium dynamical mean-field theory}
One of the simplest models to capture essential phenomena in strongly-correlated electron materials is the single-band Hubbard Hamiltonian
\begin{align}\label{eq:halffilledhub}
\hat{H}_{\mathrm{Hubbard}}=&-\sum_{\langle i, j \rangle \sigma} v_{ij}(t) \left( \hat{c}^{\dagger}_{i,\sigma} \hat{c}_{j,\sigma} + \mathrm{h.c.} \right)\nonumber \\  &+ U(t) \sum_i \left(\hat{n}_{i,\uparrow}-\frac{1}{2}\right)\left(\hat{n}_{i,\downarrow}-\frac{1}{2}\right),
\end{align}
where $v_{ij}(t)$ is the tunnelling (`hopping') matrix element between nearest-neighbour sites $i$ and $j$, and $U(t)$ is the on-site Coulomb repulsion. Here, we have assumed general time-dependent parameters due to the driving of material via, e.g., intense laser pulses~\cite{wall2011quantumSup}. Furthermore, $\hat{c}^{\dagger}_{i,\sigma}$ ($\hat{c}_{i,\sigma}$) is the creation (annihilation) operator for an electron with spin projection $\sigma=\uparrow, \downarrow$ at site $i$, while $\hat{n}_{i,\sigma}=\hat{c}^{\dagger}_{i,\sigma} \hat{c}_{i,\sigma}$ is the corresponding number operator.

Despite its apparent simplicity, the Hubbard model~\eqref{eq:halffilledhub} is notoriously difficult to solve, even numerically, and especially in two dimensions where it may be relevant to high-$T_c$ superconductivity. Fortunately, dynamical mean-field theory (DMFT)~\cite{georges1996dynamicalSup} and its extension to non-equilibrium problems~\cite{RevModPhys.86.779Sup} provide a means to compute local observables by circumventing the necessity of dealing directly with the Hubbard Hamiltonian. This is achieved by mapping it onto an impurity model, the solution of which is usually easier to obtain, albeit still a highly non-trivial computational task. The mapping is justified in the limit of infinite spatial dimensions, $d \rightarrow \infty$, (or infinite coordination, $z \rightarrow \infty$) by the collapse of the irreducible self-energy of the Hubbard model to only contributions emerging from strictly local skeleton diagrams which are identical to those of an impurity model. The collapse of the skeleton diagrams follows from the necessity to scale the hopping parameters as  $v_{ij}(t)=v^*/\sqrt{z}$ to avoid a diverging average kinetic energy per lattice site and from simple power counting arguments. While describing the full Hubbard Hamiltonian with a single-impurity model is only an approximation in finite dimensions, it often relatively accurate already in three dimensions for certain lattice types.

\begin{figure}
\centerline{
\includegraphics[scale=0.1]{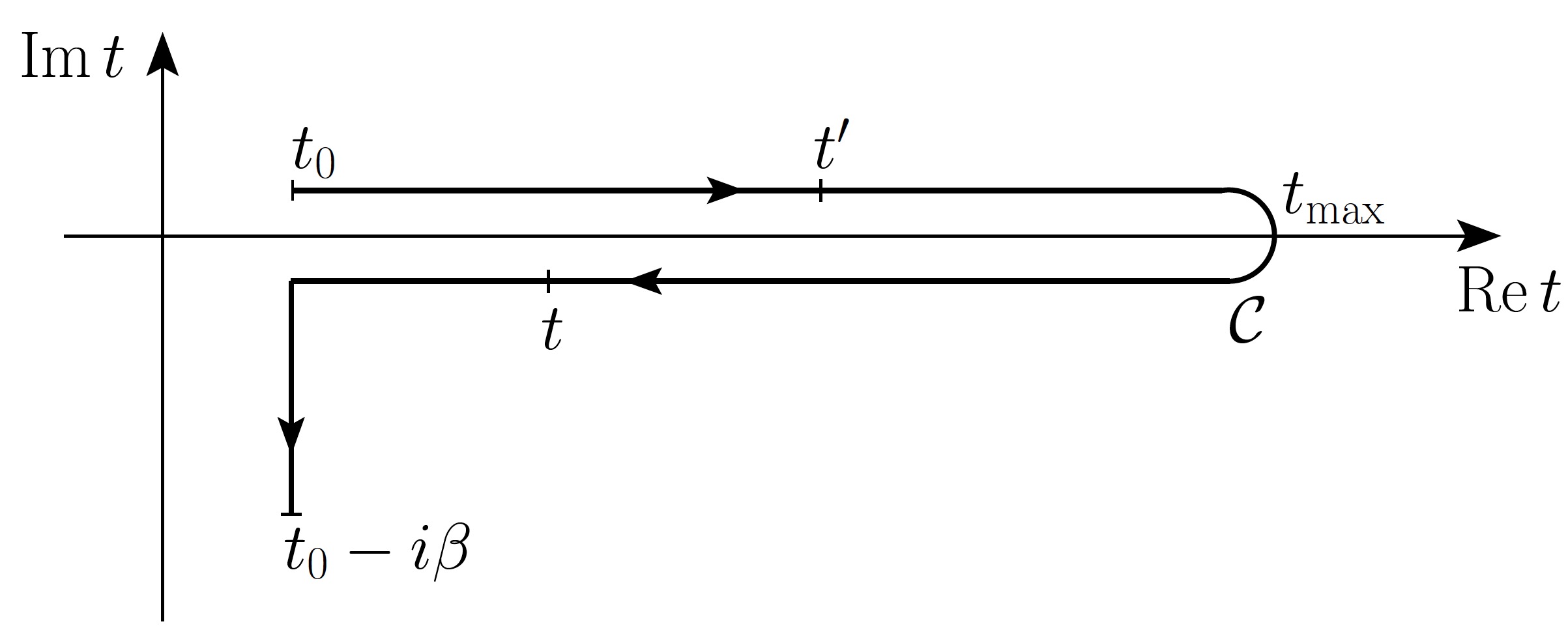}
}
\caption{Keldysh time-contour $\mathcal{C}$. It consists of two real-time branches between an initial time $t_0$ to final time $t_{\rm max}$, and an imaginary-time branch from $t_0$ to $t_0-i\beta$, where $\beta$ is the inverse temperature.}
\label{fig:keldysh}
\end{figure}

The solution of the impurity model means essentially computing the local Green function
%(for a translation-invariant system)
\begin{align}\label{eq:gf}
G_{\sigma}(t,t')&=-i\langle \hat{c}_{\sigma}(t) \hat{c}^{\dagger}_{\sigma}(t') \rangle_\mathcal{\hat{S}_{\mathrm{loc}}}\nonumber \\ &= -i \frac{\tr \left\{ \mathcal{T}_{\mathcal{C}} \left[ \exp(\hat{S}_{\mathrm{loc}})\hat{c}_{\sigma}(t) \hat{c}^{\dagger}_{\sigma}(t') \right]\right\}}{\tr \left\{ \mathcal{T}_{\mathcal{C}} \left[ \exp(\hat{S}_{\mathrm{loc}})\right]\right\} },
\end{align}
where $\mathcal{T}_{\mathcal{C}}$ is the contour-ordering operator on an `L-shaped' Keldysh time-contour $\mathcal{C}$ (see Fig.~\ref{fig:keldysh}). The local action $\hat{S}_{\mathrm{loc}}$ is given by~\cite{gramsch2013hamiltonianSup}
% in the grand-canonical ensemble 
\begin{align}\label{eq:action}
&\mathcal{\hat{S}_{\mathrm{loc}}}\nonumber \\&=-i \int_{\mathcal{C}}  dt \, \left[ U(t) \left(\hat{n}_{\uparrow}(t)-\frac{1}{2}\right)\left(\hat{n}_{\downarrow}(t)-\frac{1}{2}\right)-\mu\sum_\sigma \hat{n}_\sigma(t)\right]\nonumber \\&-i\int_{\mathcal{C}} dt \int_{\mathcal{C}}  dt' \, \sum_\sigma \Lambda_\sigma(t,t')\hat{c}^{\dagger}_\sigma(t)\hat{c}_\sigma(t').
\end{align}
Here, $\mu$ is the chemical potential and $\Lambda_\sigma$ is the \textit{a priori} unknown hybridization function, or Weiss function, that describes the exchange of electrons between the impurity site with a bath of non-interacting electrons. The essential step in DMFT is the self-consistent determination of $\Lambda_\sigma$. For a Bethe lattice, which corresponds to a semi-elliptical density of states $D(\epsilon)=\sqrt{4v^2-\epsilon^2}/(2\pi v^2)$, the DMFT self-consistency condition obtains a simple closed form. For time-dependent hoppings $v$, this reads
\begin{align}\label{eq:bethecond}
\Lambda_{\sigma}(t,t')=v(t)G_{\sigma}(t,t')v(t').
\end{align}

The impurity action~\eqref{eq:action} can also be represented in a Hamiltonian form which permits the application of Hamiltonian-based numerical methods~\cite{gramsch2013hamiltonianSup,PhysRevB.90.235131Sup} to compute the local Green function. It also makes it possible to use the trapped-ion scheme for quantum simulations~\cite{blatt2012quantumSup}. The impurity model that we address here is the single-impurity Anderson model (SIAM) given by
\begin{align}\label{eq:SIAM}
\hat{H}_{\mathrm{SIAM}}=\hat{H}_{\mathrm{loc}}+\hat{H}_{\mathrm{bath}}+\hat{H}_{\mathrm{hyb}},
\end{align}
\begin{align}
\hat{H}_{\mathrm{loc}}=-\mu \sum_\sigma \hat{n}_{\sigma} + U(t)\, \left(\hat{n}_{\uparrow}-\frac{1}{2}\right) \left(\hat{n}_{\downarrow}-\frac{1}{2}\right),
\end{align}
\begin{align}\label{eq:hhyb}
\hat{H}_{\mathrm{hyb}}=\sum_{p} \left( V_{p\sigma}(t)\hat{a}_{\sigma}^{\dagger}\hat{a}_{p\sigma} + \mathrm{H.c.} \right),
\end{align}
\begin{align}
\hat{H}_{\mathrm{bath}}=\sum_{p,\sigma} \left[ \epsilon_{p\sigma}(t)-\mu \right]\hat{a}_{p\sigma}^{\dagger}\hat{a}_{p\sigma},
\end{align}
Here, $\hat{a}_{\sigma}^{\dagger}$ ($\hat{a}_{\sigma}$) is the creation (annihilation) operator for the impurity orbital, and $\hat{a}_{p\sigma}^{\dagger}$ ($\hat{a}_{p\sigma}$) for a bath orbital $p$. Further, $V_{p\sigma}(t)$ describes the hopping of electrons between the impurity and the bath, and $\epsilon_{p\sigma}(t)$ denotes the energy of the bath orbital $p$.

The SIAM Hamiltonian~\eqref{eq:SIAM} corresponds to the correct DMFT action~\eqref{eq:action} if the parameters $V_{p\sigma}$ and $\epsilon_{p\sigma}(t)$ are chosen such that the relation
\begin{align}
\Lambda_\sigma^{\mathrm{SIAM}}(t,t')=\Lambda_\sigma(t,t')
\end{align}
is valid on the whole Keldysh contour $\mathcal{C}$. Here, the SIAM hybridization function has the expression~\cite{gramsch2013hamiltonianSup}
\begin{align}
\Lambda_\sigma^{\mathrm{SIAM}}(t,t')=\sum_p V_{p\sigma}(t)  g_{p\sigma}(t,t') V_{p\sigma}(t')^*,
\end{align}
where
\begin{align}\label{eq:qpsigma}
g_{p\sigma}(t,t') =i \left[ f(\epsilon_{p\sigma}(0)-\mu ) - \Theta_{\mathcal{C}}(t,t') \right]e^{-i\int_{\mathcal{C}}d\bar{t} \, (\epsilon_{p\sigma}(\bar{t})-\mu ) }
\end{align}
is the non-interacting Green function for an isolated bath site, with $f(\epsilon)=1/\left(\exp(\beta \epsilon) +1 \right)$ denoting the Fermi distribution function and $\Theta_{\mathcal{C}}(t,t')$ being the contour Heaviside function defined as
\begin{align}
\Theta_{\mathcal{C}}(t,t')= \begin{cases} 
      1 &  \mathrm{if} \; t \geq_{\mathcal{C}} t' \\
      0 & \mathrm{else}.
   \end{cases}
\end{align}

An essential part of the Hamiltonian-based DMFT scheme is the determination of the parameters $V_{p\sigma}(t)$ and $\epsilon_{p\sigma}(t)$ for a given hybridization function $\Lambda_\sigma(t,t')$. In what follows, we will relax the spin index $\sigma$ for the hybridization function since below we will be dealing with a spin-symmetric set-up where both contributions are identical.

For non-equilibrium problems, it is useful introduce two distinct baths, with each having their own corresponding hybridization function~\cite{gramsch2013hamiltonianSup}. The first bath, with hybridization $\Lambda_-$, includes those sites that are coupled to the impurity at $t=0$. Often this first bath vanishes as $t \rightarrow \infty$. The second bath, $\Lambda_+$, builds up as time evolves, i.e., couples additional bath sites to the impurity for times $t>0$. We will consider a system with no initial correlations ($\Lambda_-=0$) in the next section, and focus only on the second bath, with Weiss function
\begin{align}
\Lambda_+(t,t')=\sum_p V^+_{p}(t)g_{p}(t,t')V^+_{p}(t')^*.
\end{align} 
Since all imaginary-time components, which account for initial correlations, vanish for $\Lambda_+$, we set $V^+_{p}(t=0)=0$ for all bath sites that are included in $\Lambda_+$. The time-dependence of the bath energies $\epsilon_{p}(t)$ can be absorbed in the time dependence of the hoppings $V^+_{p}(t)$, meaning that we are free to choose the evolution~\cite{gramsch2013hamiltonianSup}
\begin{align}
\epsilon_p(t)= \begin{cases} 
       \epsilon_p(0) &  \mathrm{for} \; t = 0\\
      \epsilon(\infty) & \mathrm{for} \; t>0
   \end{cases},
\end{align}
where $\epsilon(\infty)$ is a constant. Moreover, since $\epsilon_p(0)$ is incorporated only in the Fermi functions $f[\pm(\epsilon_p(0)-\mu)]$ for $\Lambda^{</>}_+$, we can simply choose $\epsilon_p(0)$ such that $f(\epsilon_p(0)-\mu)$ is equal to 0 or 1. This is done in order to find a representation of the Weiss functions as
\begin{align}
-i\Lambda_+^<(t,t')=\sum_{p \in B_{\mathrm{occ}}} V^+_{p}(t)V^+_{p}(t')^*,
\end{align}
\begin{align}
i\Lambda_+^>(t,t')=\sum_{p \in B_{\mathrm{empty}}} V^+_{p}(t)V^+_{p}(t')^*,
\end{align}
in which $ B_{\mathrm{occ}}$ and $ B_{\mathrm{empty}}$ denote the sets of initially occupied and empty bath sites, respectively. Note that for a particle-hole symmetric system, $\Lambda_+^<(t,t')=\Lambda_+^>(t,t')^*$, which is satisfied if the occupied and empty bath sites come in pairs with complex conjugate hybridizations. Moreover, for a discretized time $t_n = n \times \Delta t \in [0,\, N\times \Delta t = t_{\max}]$, we have, e.g.,
\begin{align}\label{eq:chol}
(-i\Lambda_+^<)_{nn'}=-i\Lambda_+^<(t_n,t_{n'})=\sum_{p \in B_{\mathrm{occ}}} V^+_{p}(t_n)V^+_{p}(t_{n'})^*,
\end{align}
which has the form of a Cholesky decomposition $(-i\Lambda_+^<)=VV^{\dagger}$ where $V$ is a lower triangular matrix, the $p$th column of which gives the time-dependent hybridization to the bath orbital $p$. The use of Cholesky decomposition to the determine the hybridizations from the Weiss function allows us to adopt a time-propagation scheme in which we do not update the whole Green and Weiss function matrices as time evolves but only the current time slice. In practice, since we only have a limited number of bath sites $L$, we employ an approximate representation of the Weiss function in which we obtain the evolution of the first $L$ time-steps from the Cholesky decomposition, and for time steps greater than $L$ we update a new column and row in the Weiss function matrix in a manner which minimizes the error in the approximate representation~\cite{gramsch2013hamiltonianSup}. In the next section we present a test system and use the results of this section to determine the Weiss field self-consistently for the Hubbard model in an infinite-dimensional Bethe lattice.
\section{The set-up and DMFT steps}
We consider the time evolution of the infinite-dimensional Hubbard model with constant on-site interaction $U$ and time-dependent hopping $v(t)$~\cite{gramsch2013hamiltonianSup}. The hopping is turned on from the initial value $v=0$ (i.e., the atomic limit) to the final value $v=v_0=1$, which we use as the unit of energy, with the profile
\begin{align}\label{eq:quench}
v(t)= \begin{cases} 
      \frac{1}{2}\left[1-\cos(\omega_0t) \right] &  \mathrm{for} \; t  < t_q \\
      1 & \mathrm{for} \; t \geq t_q 
   \end{cases},
\end{align}
where $\omega_0=\pi/t_q$ and $t_q>0$ is a suitable quench time. In our simulations we use $t_q=0.25/v_0$. We assume a zero temperature initial state in the paramagnetic phase in the half-filled Bethe lattice. We then map the Hubbard model onto a SIAM. Since $v(t=0)=0$, $\Lambda_-$ vanishes and the hybridization function is given by $\Lambda=\Lambda_+$. Since we have a spin- and a particle-hole symmetric system, the bath is represented with pairs of initially occupied and empty sites. We take the total number of bath sites to be $L_{\mathrm{bath}}=2L$, where $L$ is the rank of the approximate representations of $-i\Lambda^<$ and $i\Lambda^>$. The initial ground state of the SIAM has an equal number of empty and doubly occupied bath sites with energies $\epsilon_{p\sigma}=0$, and singly-occupied impurity which is spin-mixed with density matrix $\rho_0 = \left( |\uparrow \rangle \langle \uparrow | +  |\downarrow \rangle \langle \downarrow | \right)/2$. To account for occupation of the impurity site, we consider two subsystems $\alpha$ and $\beta$, in which the impurity of the system $\alpha$ ($\beta$) is initially occupied by a single $\uparrow$-electron ($\downarrow$-electron). We then compute two impurity Green functions $G^{\alpha}_{{\mathrm{imp}},\sigma}$ and $G^{\beta}_{{\mathrm{imp}},\sigma}$ the average of which yields the local lattice Green function
\begin{align}\label{eq:latticegreen}
G_{{\mathrm{loc}},\sigma}(t,t')=\frac{1}{2}\left[G^{\alpha}_{{\mathrm{imp}},\sigma} +  G^{\beta}_{{\mathrm{imp}},\sigma} \right],
\end{align}
after self-consistency has been reached. Since we are considering the Hubbard model in the Bethe lattice, the DMFT self-consistency condition is given by Eq.~\eqref{eq:bethecond}.

The non-equilibrium DMFT steps to compute the single-particle lattice Green function for a maximum simulation time $t_\mathrm{max}=N \times \Delta t$ are then the following:
%are then identical to those of p.~11 of~\cite{gramsch2013hamiltonian} but the time-propagation scheme is provided by TEBD. In addition to the GF, in the time-evolution we can calculate the time-dependent double occupation $\langle d(t)  \rangle = \langle \hat{n}_{0\uparrow} \hat{n}_{0\downarrow} \rangle$ which is also averaged over the systems $\alpha$ and $\beta$. From the double occupation we can get, e.g., the interaction energy, $\langle E_{\mathrm{int}} (t)\rangle =U\left( \langle d(t)  \rangle -\frac{1}{4} \right)$~\cite{gramsch2013hamiltonian}.
\begin{itemize}
\item[0.] Choose an initial Green function $g_0$. For iteration $n=1$, initialize the hybridization function as
$\Lambda_1(t,t')=v(t)g_0(t,t')v(t'), \; \mathrm{for} \; t,t'\leq t_\mathrm{max}$,
where $v$ is the hopping in the Hubbard Hamiltonian.
% with the Bethe lattice self-consistency condition
\item[1.] Use the Cholesky decomposition for $\Lambda_n$ to obtain the hopping parameters $V_{p}(t)$ for the $n$th iteration.

\item[2.] Use exact diagonalization techniques to compute the impurity Green functions $G^s_{{\mathrm{imp}},\sigma}=\Theta_{\mathcal{C}}(t,t')G^{s,>}_{{\mathrm{imp}},\sigma}(t,t')+\Theta_{\mathcal{C}}(t',t)G^{s,<}_{{\mathrm{imp}},\sigma}(t,t')$ for $s=\alpha$ and $s=\beta$, where
\begin{eqnarray}
G^{s,>}_{{\mathrm{imp}},\sigma}(t,t')&=&-i \langle \psi_0^s | \hat{U}(0,t)\hat{c}_{1\sigma}\hat{U}(t,t')\hat{c}^{\dagger}_{1\sigma}\hat{U}(t',0)| \psi_0^s \rangle,\nonumber \\
G^{s,<}_{{\mathrm{imp}},\sigma}(t,t')&=&i \langle \psi_0^s | \hat{U}(0,t')\hat{c}^{\dagger}_{1\sigma}\hat{U}(t',t)\hat{c}_{1\sigma}\hat{U}(t,0)| \psi_0^s \rangle,\nonumber \\
\hat{U}(t,t')&=&\mathcal{T}e^{-i \int_{t'}^t \, d\tau \, \hat{H}_{\mathrm{SIAM}}(\tau)}.
\end{eqnarray}
%
%\end{align}
%\begin{align}
%G^{s,<}_{0\sigma}(t,t')=i \langle \psi_0^s | U(0,t')\hat{c}^{\dagger}_{0\sigma}U(t',t)\hat{c}_{0\sigma}U(t,0)| \psi_0^s \rangle,\nonumber
%\end{align}
%\begin{align}
%U(t,t')=\mathcal{T}e^{-i \int_{t'}^t \, d\tau \, \hat{H}_{\mathrm{imp}}(\tau)},
%\end{align}
Here, $| \psi_0^s \rangle$ is the initial (pure) state for system $s$, and $\mathcal{T}$ is the (usual) time-ordering operator. Use Eq.~\eqref{eq:latticegreen} to obtain the local lattice Green function $G_n$.

\item[3.] Use the DMFT self-consistency condition for the Bethe lattice $\Lambda_{n+1}(t,t')=v(t)G_n(t,t')v(t')$ to obtain the hybridization function for the next iteration.

\item[4.] Go to step 1 and iterate the steps until convergence is reached. The convergence variable can be, e.g., $\max | V_{p}(t)-V_{p,\mathrm{prev}}(t) |$.
\end{itemize}
From the lattice Green function we can obtain single-particle observables. In addition to the Green function, in the time-evolution we can calculate the time-dependent double occupation $\langle d(t)  \rangle = \langle \hat{n}_{1\uparrow} \hat{n}_{1\downarrow} \rangle$ which is also averaged over the systems $\alpha$ and $\beta$.

\section{Jordan--Wigner transformation applied to the single-impurity Anderson model}
The aim of the main article is to show how such DMFT steps as described above could be performed on a trapped-ion quantum computer in conjunction with a classical feedback loop. To this end, we must represent the SIAM Hamiltonian~\eqref{eq:SIAM} with $\mu=0$ and $\epsilon_{p\sigma}=0$ in terms of spin operators that operate on the qubits. This is achieved via the Jordan--Wigner transformation, in which we map a string of $N$ fermions onto a string of $2N$ qubits. The relation between the fermionic creation and annihilation operators and the spin operators reads
\begin{eqnarray}
\hat{a}^{\dagger}_{p\downarrow}&=&\hat{\sigma}_{1}^z  \otimes \cdots \otimes \hat{\sigma}_{2p-2}^z \otimes \hat{\sigma}^{-}_{2p-1} , \label{eq:jwt1}  \\ %\otimes  I^{\otimes (2N-2p+1)}
\hat{a}^{\dagger}_{p\uparrow}&=& \hat{\sigma}_{1}^z \otimes  \cdots \otimes \hat{\sigma}_{2p-1}^z \otimes \hat{\sigma}^{-}_{2p} ,\label{eq:jwt2} \\ %\otimes I^{\otimes (2N-2p)}
\hat{a}_{p\sigma}&=&(\hat{a}^{\dagger}_{p\sigma})^{\dagger},\label{eq:jwt3}
\end{eqnarray}
where $\hat{\sigma}^{\pm}=\frac{1}{2}(\hat{\sigma}^x \pm i \hat{\sigma}^y)$, and $\hat{\sigma}^x$, $\hat{\sigma}^y$, and $\hat{\sigma}^z$ are the Pauli spin operators.

We apply the transformations \eqref{eq:jwt1}-\eqref{eq:jwt3} to the SIAM Hamiltonian. The interaction term becomes
\begin{align}
U(t)\left(\hat{n}_\uparrow - \frac{1}{2} \right)\left(\hat{n}_\downarrow - \frac{1}{2} \right) = \frac{1}{4}U(t) \hat{\sigma}^z_1 \otimes \hat{\sigma}_2^z ,
\end{align}
while the hybridization terms read
\begin{widetext}
\begin{align}\label{eq:hybdn}
V_{p\downarrow}\hat{a}_{1\downarrow}^{\dagger}\hat{a}_{p\downarrow}+\mathrm{H.c.}=&\; \frac{1}{2}\mathrm{Re}(V_{p\downarrow})\left( \hat{\sigma}_1^x \otimes \hat{\sigma}_2^z \otimes \cdots \otimes \hat{\sigma}_{2p-2}^z \otimes \hat{\sigma}_{2p-1}^x + \hat{\sigma}_1^y \otimes \hat{\sigma}_2^z \otimes \cdots \otimes \hat{\sigma}_{2p-2}^z \otimes \hat{\sigma}_{2p-1}^y \right) \nonumber \\
 &+\frac{1}{2}\mathrm{Im}(V_{p\downarrow})\left( \hat{\sigma}_1^y \otimes \hat{\sigma}_2^z \otimes \cdots \otimes \hat{\sigma}_{2p-2}^z \otimes \hat{\sigma}_{2p-1}^x - \hat{\sigma}_1^x \otimes \hat{\sigma}_2^z \otimes \cdots \otimes \hat{\sigma}_{2p-2}^z \otimes \hat{\sigma}_{2p-1}^y \right),
\end{align}

\begin{align}\label{eq:hybup}
V_{p\uparrow}\hat{a}_{1\uparrow}^{\dagger}\hat{a}_{p\uparrow}+\mathrm{H.c.}=&\; \frac{1}{2}\mathrm{Re}(V_{p\uparrow})\left( \hat{\sigma}_2^x \otimes \hat{\sigma}_3^z \otimes \cdots \otimes \hat{\sigma}_{2p-1}^z \otimes \hat{\sigma}_{2p}^x + \hat{\sigma}_2^y \otimes \hat{\sigma}_3^z \otimes \cdots \otimes \hat{\sigma}_{2p-1}^z \otimes \hat{\sigma}_{2p}^y \right) \nonumber \\
 &+\frac{1}{2}\mathrm{Im}(V_{p\uparrow})\left( \hat{\sigma}_2^y \otimes \hat{\sigma}_3^z \otimes \cdots \otimes \hat{\sigma}_{2p-1}^z \otimes \hat{\sigma}_{2p}^x - \hat{\sigma}_2^x \otimes \hat{\sigma}_3^z \otimes \cdots \otimes \hat{\sigma}_{2p-1}^z \otimes \hat{\sigma}_{2p}^y \right).
\end{align}
\end{widetext}
In order to implement the time-evolution operator in an experiment, we use the Trotter decomposition
\begin{align}
e^{-i\delta t\sum_{j=1}^N \hat{h}_j} \approx \prod_{j=1}^N e^{-i\delta t \hat{h}_j},
\end{align}
in which each of the terms on the right hand side can be implemented with the help of M{\o}lmer--S{\o}rensen gates and local and global rotations, as described in the next section.
\section{Implementing the SIAM Hamiltonian with M{\o}lmer--S{\o}rensen gates}
Each exponent that consists of tensor products of $k$ Pauli operators can be implemented (up to local rotations) with a  M{\o}lmer--S{\o}rensen gate acting on the $k$ qubits, one local gate acting on a single qubit, and the inverse  M{\o}lmer--S{\o}rensen gate~\cite{casanova2012quantumSup,muller2011simulatingSup}. For example, we have the decomposition
\begin{align}\label{eq:MSlocMSx}
\hat{U}&=\hat{U}_{\mathrm{MS}}^{1,k}\left(-\frac{\pi}{2},0\right)\hat{U}_{1,\mathrm{loc}}(\phi)\hat{U}_{\mathrm{MS}}^{1,k}\left(\frac{\pi}{2},0\right)\nonumber \\
&=\exp\left({i\phi \sigma_1^z \otimes \sigma_2^x \otimes \sigma_3^x \otimes \cdots \otimes \sigma_k^x}\right),
\end{align}
where the M{\o}lmer--S{\o}rensen gate is given by 
\begin{align}
\hat{U}_{\mathrm{MS}}^{l,m}\left( \theta, \phi \right)=\exp \left[ -i\frac{\theta}{4}\left(\cos\phi \, \hat{S}_x + \sin\phi \hat{S}_y \right)^2 \right],
\end{align}
with $\hat{S}_{x,y}=\sum_{j=l}^m \hat{\sigma}_j^{x,y}$. The local gate in Eq.~\eqref{eq:MSlocMSx} reads
\begin{align}
\hat{U}_{j,\mathrm{loc}}(\phi)=\begin{cases} 
      \exp(-i\phi \sigma_j^z) &  \mathrm{for} \; k=4n-1 \\
      \exp(i\phi \sigma_j^z) &  \mathrm{for} \; k=4n+1 \\
      \exp(-i\phi \sigma_j^y) &  \mathrm{for} \; k=4n-2 \\
      \exp(i\phi \sigma_j^y) &  \mathrm{for} \; k=4n \\
   \end{cases},\quad n\in \mathbb{N},
\end{align}
To implement a string of $\hat{\sigma}^y$ gates instead of $\hat{\sigma}^x$, we use a different M{\o}lmer--S{\o}rensen gate, yielding the decomposition
\begin{align}\label{eq:MSlocMSy}
\hat{U}&=\hat{U}_{\mathrm{MS}}^{1,k}\left(-\frac{\pi}{2},\frac{\pi}{2}\right)\hat{U}_{1,\mathrm{loc}}(\phi)\hat{U}_{\mathrm{MS}}^{1,k}\left(\frac{\pi}{2},\frac{\pi}{2}\right)\nonumber \\
&=\exp\left({i\phi \sigma_1^z \otimes \sigma_2^y \otimes \sigma_3^y \otimes \cdots \otimes \sigma_k^y}\right),
\end{align}
with the local gate
\begin{align}
\hat{U}_{j,\mathrm{loc}}(\phi)=\begin{cases} 
      \exp(-i\phi \sigma_j^z) &  \mathrm{for} \; k=4n-1 \\
      \exp(i\phi \sigma_j^z) &  \mathrm{for} \; k=4n+1 \\
      \exp(i\phi \sigma_j^x) &  \mathrm{for} \; k=4n-2 \\
      \exp(-i\phi \sigma_j^x) &  \mathrm{for} \; k=4n \\
   \end{cases},\quad n\in \mathbb{N}.
\end{align}
Any of the gates from Eqs.~\eqref{eq:hybdn} and~\eqref{eq:hybup} can be obtained from Eqs.~\eqref{eq:MSlocMSx} and~\eqref{eq:MSlocMSy} by applying additional local rotations. For instance,
\begin{widetext}
\begin{align}
\exp\left(i\phi \hat{\sigma}_2^x \otimes \hat{\sigma}_3^z \otimes \cdots \otimes \hat{\sigma}_{2p-1}^z \otimes \hat{\sigma}_{2p}^x \right)=\exp\left(i \frac{\pi}{4}\sum_{j=4}^{2p-1} \hat{\sigma}_j^y \right)\hat{U}_{\mathrm{MS}}^{2,2p}\left(-\frac{\pi}{2},0\right)\hat{U}_{3,\mathrm{loc}}(\phi)\hat{U}_{\mathrm{MS}}^{2,2p}\left(\frac{\pi}{2},0\right)\exp\left(-i \frac{\pi}{4}\sum_{j=4}^{2p-1} \hat{\sigma}_j^y \right),
\end{align}
where $\hat{U}_{3,\mathrm{loc}}(\phi)=\exp\left(-i\phi \hat{\sigma}_3^z \right)$ for even $p$, and $\hat{U}_{3,\mathrm{loc}}(\phi)=\exp\left(i\phi \hat{\sigma}_3^z \right)$ for odd $p$, with $\phi = -\frac{1}{2}\delta t \mathrm{Re}(V_{p\uparrow})$. Similarly, e.g.,
\begin{align}
&\exp\left(i\phi \hat{\sigma}_1^x \otimes \hat{\sigma}_2^z \otimes \cdots \otimes \hat{\sigma}_{2p-2}^z \otimes \hat{\sigma}_{2p-1}^y \right)\nonumber \\ &=\exp\left( i\frac{\pi}{4}\hat{\sigma}_1^z \right) \exp\left(-i \frac{\pi}{4}\sum_{j=3}^{2p-2} \hat{\sigma}_j^y \right)\hat{U}_{\mathrm{MS}}^{1,2p-1}\left(-\frac{\pi}{2},\frac{\pi}{2}\right)\hat{U}_{2,\mathrm{loc}}(\phi)\hat{U}_{\mathrm{MS}}^{1,2p-1}\left(\frac{\pi}{2},\frac{\pi}{2}\right)\exp\left(i \frac{\pi}{4}\sum_{j=3}^{2p-2} \hat{\sigma}_j^y \right)\exp\left( -i\frac{\pi}{4}\hat{\sigma}_1^z \right),
\end{align}
where $\hat{U}_{2,\mathrm{loc}}(\phi)=\exp\left(-i\phi \hat{\sigma}_2^z \right)$ for even $p$, and $\hat{U}_{2,\mathrm{loc}}(\phi)=\exp\left(i\phi \hat{\sigma}_2^z \right)$ for odd $p$, with $\phi = \frac{1}{2}\delta t \mathrm{Im}(V_{p\downarrow})$.
\end{widetext}

\section{Measuring the local Green function}
An essential part of the scheme is the determination of the local non-equilibrium Green function. In this section, we propose an experimental scheme to measure it with trapped ions. We again apply the Jordan--Wigner transformations on the $\hat{c}$-operators and obtain the following expressions for the different components of the Green function
\begin{widetext}
\begin{align}\label{eq:gsgup}
G^{s,>}_{1\uparrow}(t,t')=&-\frac{i}{4}  \Big (\langle \psi_0^s | \hat{U}(0,t)( \hat{\sigma}^z_1 \otimes \hat{\sigma}^x_2) \hat{U}(t,t')  ( \hat{\sigma}^z_1 \otimes \hat{\sigma}^x_2) \hat{U}(t',0)| \psi_0^s \rangle -i \langle \psi_0^s | \hat{U}(0,t)( \hat{\sigma}^z_1 \otimes \hat{\sigma}^x_2) \hat{U}(t,t')  (\hat{\sigma}^z_1  \otimes \hat{\sigma}^y_2) \hat{U}(t',0)| \psi_0^s \rangle \nonumber \\ & + i\langle \psi_0^s | \hat{U}(0,t)( \hat{\sigma}^z_1 \otimes \hat{\sigma}^y_2) \hat{U}(t,t')  ( \hat{\sigma}^z_1 \otimes \hat{\sigma}^x_2) \hat{U}(t',0)| \psi_0^s \rangle  +\langle \psi_0^s | \hat{U}(0,t)( \hat{\sigma}^z_1 \otimes \hat{\sigma}^y_2) \hat{U}(t,t')  ( \hat{\sigma}^z_1 \otimes \hat{\sigma}^y_2) \hat{U}(t',0)| \psi_0^s \rangle \Big ),
\end{align}
\begin{align}\label{eq:gsgdn}
G^{s,>}_{1\downarrow}(t,t')=&-\frac{i}{4}  \Big ( \langle \psi_0^s | \hat{U}(0,t)\hat{\sigma}^x_1 \hat{U}(t,t')  \hat{\sigma}^x_1 \hat{U}(t',0)| \psi_0^s \rangle -i \langle \psi_0^s | \hat{U}(0,t)\hat{\sigma}^x_1  \hat{U}(t,t')  \hat{\sigma}^y_1  \hat{U}(t',0)| \psi_0^s \rangle \nonumber \\ & +i \langle \psi_0^s | \hat{U}(0,t)\hat{\sigma}^y_1  \hat{U}(t,t')  \hat{\sigma}^x_1  \hat{U}(t',0)| \psi_0^s \rangle  +\langle \psi_0^s | \hat{U}(0,t)\hat{\sigma}^y_1  \hat{U}(t,t') \hat{\sigma}^y_1  \hat{U}(t',0)| \psi_0^s \rangle \Big ),
\end{align}
\begin{align}\label{eq:gslup}
G^{s,<}_{1\uparrow}(t,t')=&\frac{i}{4}  \Big ( \langle \psi_0^s | \hat{U}(0,t')(\hat{\sigma}^z_1  \otimes \hat{\sigma}^x_2) \hat{U}(t',t)  ( \hat{\sigma}^z_1 \otimes \hat{\sigma}^x_2) \hat{U}(t,0)| \psi_0^s \rangle +i \langle \psi_0^s | \hat{U}(0,t')(\hat{\sigma}^z_1 \otimes  \hat{\sigma}^x_2 ) \hat{U}(t',t)  (\hat{\sigma}^z_1  \otimes \hat{\sigma}^y_2) \hat{U}(t,0)| \psi_0^s \rangle \nonumber \\ & -i \langle \psi_0^s | \hat{U}(0,t')(\hat{\sigma}^z_1  \otimes \hat{\sigma}^y_2) \hat{U}(t',t)  (\hat{\sigma}^z_1  \otimes \hat{\sigma}^x_2) \hat{U}(t,0)| \psi_0^s \rangle  + \langle \psi_0^s | \hat{U}(0,t')(\hat{\sigma}^z_1  \otimes \hat{\sigma}^y_2) \hat{U}(t',t)  (\hat{\sigma}^z_1  \otimes \hat{\sigma}^y_2) \hat{U}(t,0)| \psi_0^s \rangle \Big ),
\end{align}
\begin{align}\label{eq:gsldn}
G^{s,<}_{1\downarrow}(t,t')=&\frac{i}{4}  \Big ( \langle \psi_0^s | \hat{U}(0,t')\hat{\sigma}^x_1 \hat{U}(t',t)  \hat{\sigma}^x_1 \hat{U}(t,0)| \psi_0^s \rangle +i \langle \psi_0^s | \hat{U}(0,t')\hat{\sigma}^x_1  \hat{U}(t',t)  \hat{\sigma}^y_1  \hat{U}(t,0)| \psi_0^s \rangle \nonumber \\ & -i \langle \psi_0^s | \hat{U}(0,t')\hat{\sigma}^y_1  \hat{U}(t',t)  \hat{\sigma}^x_1  \hat{U}(t,0)| \psi_0^s \rangle  + \langle \psi_0^s | \hat{U}(0,t')\hat{\sigma}^y_1  \hat{U}(t',t) \hat{\sigma}^y_1  \hat{U}(t,0)| \psi_0^s \rangle \Big ),
\end{align}
\end{widetext}

In Eqs.~\eqref{eq:gsgup}-\eqref{eq:gsldn}, all time-evolution operators $\hat{U}(t,0)$, etc, correspond to a sequence of quantum gates obtained in the previous section. 

To measure each of the summands in Eqs.~\eqref{eq:gsgup}-\eqref{eq:gsldn},
we introduce a probe qubit~\cite{dorner2013extractingSup} which we couple to the system of interest. We assume that the probe qubit is prepared in the pure state $|0\rangle$, yielding the total density operator $\hat{\rho}_{\mathrm{tot}}=\hat{\rho}_{\mathrm{sys}}\otimes |0\rangle \langle 0|$. The combined system is then run through a Ramsey interferometer sequence described by a quantum circuit in which we first apply a Hadamard gate $\hat{\sigma}_H$ ($\pi/2$ pulse) on the probe qubit, followed by unitary evolution of the system of interest, followed by a controlled application of Pauli gates, evolution up to the final time, another controlled application of Pauli gates, and ending with another Hadamard gate on the probe qubit (see Fig.~2 of the main text). The output state of the qubit at the end of the Ramsey sequence is given by %(THE MINUS SIGNS IN THE OFF-DIAGONAL ELEMENTS MAY BE THE WRONG WAY AROUND --- CHECK!)
\begin{align}
\hat{\rho}_{\mathrm{probe}}=&\tr_{\mathrm{sys}}\left[\hat{\sigma}_H \hat{T} \hat{\sigma}_H \hat{\rho}_{\mathrm{tot}} \hat{\sigma}_H \hat{T}^{\dagger} \hat{\sigma}_H \right]\nonumber \\
=&\frac{1+\mathrm{Re}[F(t,t')]}{2}|0\rangle \langle 0 | - i\frac{ \mathrm{Im}[F(t,t')]}{2}|0 \rangle \langle 1 | \nonumber \\ & + i\frac{ \mathrm{Im}[F(t,t')]}{2}|1 \rangle \langle 0 | + \frac{1-\mathrm{Re}[F(t,t')]}{2}|1 \rangle \langle 1 |,
\end{align}
where $F(t,t')=\tr_{\mathrm{sys}}\left[ \hat{T}_{1}^{\dagger}(t) \hat{T}_{0}(t,t') \hat{\rho}_{\mathrm{sys}} \right]$ 
corresponds to one of the summands in Eqs.~\eqref{eq:gsgup}-\eqref{eq:gsldn}. Here, the unitary operators $\hat{T}_{0}(t,t')=\langle 0 | \hat{T} | 0 \rangle=\hat{U}(t,t')  \hat{\sigma} \hat{U}(t',0)$ and $\hat{T}_{1}(t)=\langle 1 | \hat{T} | 1 \rangle = \hat{{\sigma}}' \hat{U}(t,0)$,  in which $\hat{\sigma}$ and $\hat{{\sigma}}'$ are Pauli operators or tensor products of Pauli operators according to Eqs.~\eqref{eq:gsgup}-\eqref{eq:gsldn}, act only on the system and not on the probe qubit. For example, the network in Fig.~2 of the main text corresponds to the case $\hat{\sigma}=\hat{\sigma}^z_1 \otimes \hat{\sigma}^x_2$ and $\hat{\sigma}'=\hat{\sigma}^z_1 \otimes \hat{\sigma}^y_2$. Note that
\begin{align}
\hat{\rho}_{\mathrm{probe}}=\frac{1}{2}\left(\hat{I} + \mathrm{Re}[F(t,t')]\hat{\sigma}_z +\mathrm{Im}[F(t,t')] \hat{\sigma}_y \right),
\end{align}
where $\hat{I}$ is the identity operator, so that we have
\begin{align}
\tr_{\mathrm{probe}}\left[\hat{\rho}_{\mathrm{probe}} \hat{\sigma}_z \right]=\mathrm{Re}[F(t,t')],
\end{align}
and
\begin{align}
\tr_{\mathrm{probe}}\left[\hat{\rho}_{\mathrm{probe}} \hat{\sigma}_y \right]=\mathrm{Im}[F(t,t')],
\end{align}
which are then experimentally measurable quantities. 

To give a rough estimate on the number of measurements required in an experiment, we consider the probe to be in the superposition $\frac{1}{\sqrt{2}} \left(|0\rangle + |1\rangle \right)$ which is the state with maximal uncertainty in the measurement outcome. Thus, the measurement of the $\sigma^z$ component yields either -1 or +1 with probability $\frac{1}{2}$. This random variable then follows a two-point distribution with parameters $p=0.5$, $q=1-p=0.5$, and variance $\sigma^2=1$. The mean of the $\sigma^z$ component in this state is zero. To obtain this mean with a standard error of the mean $\epsilon=\sigma / \sqrt{n}$ requires $n=\sigma^2 / \epsilon^2$ projective measurements for each contribution to the Green function. For example for $\epsilon=0.02$ we would need about  $2 \times 2 \times 2 \times 4 \times 2 \times 2\,500=160\,000$ [2 systems ($\alpha$ and $\beta$), 2 spins, lesser and greater Green function, 4 terms per Green function, 2 expectation values to be measured, and 2\,500 measurements for each expectation value] measurements per time step, and this number scales quadratically with the number of points in the time grid. However, if we consider a spin-symmetric system as above, where we have the symmetries $G_{1\sigma}^{\alpha(\beta),</>}=G_{1\bar{\sigma}}^{\beta(\alpha),</>}$, we only need to measure half of the Green functions above. Note that all measurements can be done in parallel.

\section{Outline of the classical simulations of the hybrid device}
We perform classical simulations of the single-qubit interferometer described in the previous section. In the actual hybrid device, the single-qubit interferometry would be done experimentally, and here we try to mimic the experimental procedure. 

We consider the first $L$ time steps, where $L$ is the half the number of bath sites. We first obtain some initial guess hybridization parameters $V_{p\sigma}^{(0)}(t)$, where $t=0,\Delta t,\dots, L\Delta$. Using $V_{p\sigma}^{(0)}(t)$ we construct \emph{imperfect} quantum gates $\hat{U}_{\mathrm{rot}}(\varphi+\epsilon)$ and $\hat{U}_{\mathrm{MS}}^{l,m}\left( \theta + \epsilon_{\mathrm{MS}1}, \phi+ \epsilon_{\mathrm{MS}2} \right)$, where $\epsilon$, $\epsilon_{\mathrm{MS}1}$, and $\epsilon_{\mathrm{MS}2}$ are normally distributed random variables with zero mean and standard deviations $\sigma$, $\sigma_{\rm MS1}$, and $\sigma_{\rm MS2}$, respectively. These quantum gates yield the Trotterized unitary evolution operator $\hat{U}(t,t')$, where $t,t'=0,\Delta t,\dots, L\Delta$. We use this evolution operator to compute the $(t=m\Delta t,t'=n\Delta t)$-point ($m, n \leq L$) of $F(t,t')$ from $\tr_{\mathrm{probe}}\left[\hat{\rho}_{\mathrm{probe}} \hat{\sigma}_z \right]$ and $\tr_{\mathrm{probe}}\left[\hat{\rho}_{\mathrm{probe}} \hat{\sigma}_y \right]$ as explained in the previous section, and we average the results over several realizations to gather error statistics. After going through all the possible combinations of the controlled $\hat{\sigma}^x_1$ ($\hat{\sigma}^z_1 \otimes \hat{\sigma}^x_2$) and $\hat{\sigma}^y_1$ ($\hat{\sigma}^z_1 \otimes \hat{\sigma}^y_2$) gates according to Eqs.~\eqref{eq:gsgup}-\eqref{eq:gsldn}, we obtain $G_{\downarrow (\uparrow)}(t=m\Delta t,t'=n\Delta t)$. 

However, we interpret the computation of the point $(t=m\Delta t,t'=n\Delta t)$ as a measurement which collapses the state of the system, and we cannot store any information of the state at this time instant in memory, since we don't want to re-use any of the obtained wave functions later to avoid correlating the errors between different points in the Green function. We compute these points from independent realizations instead. This way we make our classical simulations to follow what one would do in an experiment. This means that in order to compute another point $(t=(m+1)\Delta t,t'=n\Delta t)$ or $(t=m\Delta t,t'=(n+1)\Delta t)$, we have to propagate again from the origin $(t=0,t'=0)$ to the desired point and again average over several realizations. This procedure is repeated until we have obtained all the points of $G_\sigma(t,t')$ until $(t=L\Delta t,t'=L\Delta t)$. This concludes the `experimental', or quantum, part of the first $L$ time steps in the \emph{first} iteration of the DMFT self-consistency loop.

The obtained $G_\sigma(t,t')$ is then used in the classical computer to produce the hybridization function $\Lambda_{\sigma}(t,t')=v(t)G_\sigma(t,t')v(t')$. In the first $L$ time steps, we have enough parameters to do a Cholesky decomposition of $\Lambda_{\sigma}(t,t')$ to obtain new hybridizations $V_{p\sigma}^{(1)}(t)$, which are used for updating $\hat{U}_{\mathrm{rot}}(\varphi+\epsilon)$ and $\hat{U}_{\mathrm{MS}}^{l,m}\left( \theta + \epsilon_{\mathrm{MS}1}, \phi+ \epsilon_{\mathrm{MS}2} \right)$. This begins the \emph{second} iteration of the DMFT self-consistency loop where use the updated quantum gates to again `measure' $G_\sigma(t,t')$ using the steps described above, always starting from the origin to compute one point in the time grid and averaging over several realizations. This non-linear process of `measuring' $G_\sigma(t,t')$ and using Cholesky decomposition of $\Lambda_\sigma(t,t')$ to update $V_{p\sigma}(t)$ is repeated until $|V_{p\sigma}^{(n)}(t) - V_{p\sigma}^{(n-1)}(t) | < \delta$ where $\delta$ is a predetermined error threshold.

For the time steps $L+1,\dots ,N$ with $t_{\rm max}=N\Delta t$, we adopt the `time slicing' scheme of Ref.~\cite{gramsch2013hamiltonianSup}, where we iterate one time step $M>L$ to self-consistency before moving to $M+1$. In the classical part of the hybrid device, we utilize a simple minimizer step~\cite{gramsch2013hamiltonianSup} to update only $V_{p\sigma}(M\Delta t)$ while keeping the previously obtained $V_{p\sigma}(K\Delta t)$ ($K<M$) fixed. However, again when we want to reach the $M$th time step in the time grid, we have to start propagating from the origin.

Mimicking the experiment to this level makes our classical simulation very difficult.
Thus, our simulations are limited to small system sizes and relatively short time scales.

\section{Non-interacting system and error correction}
The non-interacting impurity system comprises of SIAM Hamiltonian~\eqref{eq:SIAM} with $U=0$.
We take each bath site as being independently coupled to a thermal reservoir to which it can incoherently exchange electrons with. This is described within the quantum master equation approach where the density operator $\hat{\rho}(t)$ of the full system obeys
\begin{align*}
\frac{{\rm d}}{{\rm  d}t}\hat{\rho}(t) =& -i[\hat{H}_{\rm SIAM},\hat{\rho}(t)] \\
& + \sum_{p>0,\sigma}\Gamma^-_p[2\hat{c}_{p\sigma}\hat{\rho}(t)\hat{c}^\dagger_{p\sigma} - \hat{\rho}(t)\hat{c}^\dagger_{p\sigma}\hat{c}_{p\sigma} -  \hat{c}^\dagger_{p\sigma}\hat{c}_{p\sigma}\hat{\rho}(t)] \\
& + \sum_{p>0,\sigma}\Gamma^+_p[2\hat{c}^\dagger_{p\sigma}\hat{\rho}(t)\hat{c}_{p\sigma} -  \hat{\rho}(t)\hat{c}_{p\sigma}\hat{c}^\dagger_{p\sigma} -  \hat{c}_{p\sigma}\hat{c}^\dagger_{p\sigma}\hat{\rho}(t)],
\end{align*}
where $\Gamma_p^\pm$ are the rates of electron ejection ($-$) and injection ($+$) to bath site $p$. In the case of no impurity coupling $V_{p\sigma}(t) = 0$ the noise on each bath site will drive their occupancies to a steady-state value of $n_p(\infty) = \Gamma^+_p/(\Gamma^-_p + \Gamma^+_p)$. 

Since this model is non-interacting and has Lindblad noise terms which are linear in the electron creation and annihilation operators the master equation can be solved exactly using the so-called super-fermion formalism \cite{superfermionSup}. Here we use this approach to compute the impurity single-particle Green functions
\begin{eqnarray*}
G^>_\sigma(t,t') &=& i{\rm Tr}[\hat{\rho}_0\hat{c}^\dagger_{1\sigma}(t')\hat{c}_{1\sigma}(t)], \\
G^<_\sigma(t,t') &=& -i{\rm Tr}[\hat{\rho}_0\hat{c}_{1\sigma}(t)\hat{c}^\dagger_{1\sigma}(t')],
\end{eqnarray*}
for this system given an initial density operator $\hat{\rho}_0$. We focused on a quench of the Hubbard hopping parameter $v(t)$ given by Eq.~\eqref{eq:quench}.
The initial density operator $\hat{\rho}_0$ was again chosen to model a $T=0$ half-filled paramagnetic phase \cite{gramsch2013hamiltonianSup}, where $\mu = 0$, with the impurity being in a singly occupied spin-mixed state $\half(\ket{\uparrow}\bra{\uparrow} + \ket{\downarrow}\bra{\downarrow})$, along with half the bath sites were doubly occupied $\ket{\uparrow\downarrow}$, and the other half empty $\ket{0}$. The dissipation in the bath was taken to have $\Gamma^\pm_p = \Gamma$ so that the steady-state density of the system remains a constant unit-filling. We take the bath energies to $\epsilon_{p\sigma}(t) = 0$ throughout.

Using the calculated $G^{> \atop <}_\sigma(t,t')$ the non-equilibrium DMFT self-consistency loop was solved using (i) the standard Cholesky time-slicing proposed for a noiseless system \cite{gramsch2013hamiltonianSup}, explained after Eq.~\eqref{eq:chol}, and (ii) using a fitting procedure which attempts to correct for the effects of the bath noise. We solve numerically for the bath Green functions $g_{p\sigma}(t,t')$ using the super-fermion approach~\cite{superfermionSup}. To implement a noise-reduction scheme, we minimize $\left \| \sum_p V_{p\sigma}(t)g_{p\sigma}(t,t')V_{p\sigma}(t')- \Lambda_{\sigma}(t,t') \right \|_F^2$ ($\| \cdot \|_F$ is the Frobenius norm) over the $V_{p\sigma}(t)$ to obtain the hybridizations corresponding to a noisy system. It is often useful to include a multiplying function of the form $f(t,t')=\exp(-\mu |t-t'|)$ in the cost function to aid the convergence of the minimiser.


\begin{thebibliography}{10}
\expandafter\ifx\csname url\endcsname\relax
  \def\url#1{\texttt{#1}}\fi
\expandafter\ifx\csname urlprefix\endcsname\relax\def\urlprefix{URL }\fi
\providecommand{\bibinfo}[2]{#2}
\providecommand{\eprint}[2][]{\url{#2}}

\bibitem{nqit}
\bibinfo{title}{A bet on quantum}.
\newblock \emph{\bibinfo{journal}{Nature Phys.}} \textbf{\bibinfo{volume}{11}},
  \bibinfo{pages}{89} (\bibinfo{year}{2015}).

\bibitem{barends2015digital}
\bibinfo{author}{Barends, R.} \emph{et~al.}
\newblock \bibinfo{title}{Digital quantum simulation of fermionic models with a
  superconducting circuit}.
\newblock \emph{\bibinfo{journal}{Nature Comm.}} \textbf{\bibinfo{volume}{6}},
  \bibinfo{pages}{7654} (\bibinfo{year}{2015}).

\bibitem{feynman1982simulating}
\bibinfo{author}{Feynman, R.~P.}
\newblock \bibinfo{title}{Simulating physics with computers}.
\newblock \emph{\bibinfo{journal}{Int. J. Theor. Phys.}}
  \textbf{\bibinfo{volume}{21}}, \bibinfo{pages}{467--488}
  (\bibinfo{year}{1982}).

\bibitem{buluta2009quantum}
\bibinfo{author}{Buluta, I.} \& \bibinfo{author}{Nori, F.}
\newblock \bibinfo{title}{Quantum simulators}.
\newblock \emph{\bibinfo{journal}{Science}} \textbf{\bibinfo{volume}{326}},
  \bibinfo{pages}{108--111} (\bibinfo{year}{2009}).

\bibitem{johnson2014quantum}
\bibinfo{author}{Johnson, T.~H.}, \bibinfo{author}{Clark, S.~R.} \&
  \bibinfo{author}{Jaksch, D.}
\newblock \bibinfo{title}{What is a quantum simulator?}
\newblock \emph{\bibinfo{journal}{EPJ Quantum Technology}}
  \textbf{\bibinfo{volume}{1}}, \bibinfo{pages}{1--12} (\bibinfo{year}{2014}).

\bibitem{blatt2012quantum}
\bibinfo{author}{Blatt, R.} \& \bibinfo{author}{Roos, C.~F.}
\newblock \bibinfo{title}{Quantum simulations with trapped ions}.
\newblock \emph{\bibinfo{journal}{Nature Phys.}} \textbf{\bibinfo{volume}{8}},
  \bibinfo{pages}{277--284} (\bibinfo{year}{2012}).

\bibitem{bloch2012quantum}
\bibinfo{author}{Bloch, I.}, \bibinfo{author}{Dalibard, J.} \&
  \bibinfo{author}{Nascimb{\`e}ne, S.}
\newblock \bibinfo{title}{Quantum simulations with ultracold quantum gases}.
\newblock \emph{\bibinfo{journal}{Nature Phys.}} \textbf{\bibinfo{volume}{8}},
  \bibinfo{pages}{267--276} (\bibinfo{year}{2012}).

\bibitem{houck2012chip}
\bibinfo{author}{Houck, A.~A.}, \bibinfo{author}{T{\"u}reci, H.~E.} \&
  \bibinfo{author}{Koch, J.}
\newblock \bibinfo{title}{On-chip quantum simulation with superconducting
  circuits}.
\newblock \emph{\bibinfo{journal}{Nature Phys.}} \textbf{\bibinfo{volume}{8}},
  \bibinfo{pages}{292--299} (\bibinfo{year}{2012}).

\bibitem{PhysRevA.92.062318}
\bibinfo{author}{Wecker, D.} \emph{et~al.}
\newblock \bibinfo{title}{Solving strongly correlated electron models on a
  quantum computer}.
\newblock \emph{\bibinfo{journal}{Phys. Rev. A}} \textbf{\bibinfo{volume}{92}},
  \bibinfo{pages}{062318} (\bibinfo{year}{2015}).

\bibitem{PhysRevA.93.032303}
\bibinfo{author}{Dallaire-Demers, P.-L.} \& \bibinfo{author}{Wilhelm, F.~K.}
\newblock \bibinfo{title}{Method to efficiently simulate the thermodynamic
  properties of the {Fermi-Hubbard} model on a quantum computer}.
\newblock \emph{\bibinfo{journal}{Phys. Rev. A}} \textbf{\bibinfo{volume}{93}},
  \bibinfo{pages}{032303} (\bibinfo{year}{2016}).

\bibitem{georges1996dynamical}
\bibinfo{author}{Georges, A.}, \bibinfo{author}{Kotliar, G.},
  \bibinfo{author}{Krauth, W.} \& \bibinfo{author}{Rozenberg, M.~J.}
\newblock \bibinfo{title}{Dynamical mean-field theory of strongly correlated
  fermion systems and the limit of infinite dimensions}.
\newblock \emph{\bibinfo{journal}{Rev. Mod. Phys.}}
  \textbf{\bibinfo{volume}{68}}, \bibinfo{pages}{13} (\bibinfo{year}{1996}).

\bibitem{RevModPhys.86.779}
\bibinfo{author}{Aoki, H.} \emph{et~al.}
\newblock \bibinfo{title}{Nonequilibrium dynamical mean-field theory and its
  applications}.
\newblock \emph{\bibinfo{journal}{Rev. Mod. Phys.}}
  \textbf{\bibinfo{volume}{86}}, \bibinfo{pages}{779--837}
  (\bibinfo{year}{2014}).

\bibitem{benhelm2008towards}
\bibinfo{author}{Benhelm, J.}, \bibinfo{author}{Kirchmair, G.},
  \bibinfo{author}{Roos, C.~F.} \& \bibinfo{author}{Blatt, R.}
\newblock \bibinfo{title}{Towards fault-tolerant quantum computing with trapped
  ions}.
\newblock \emph{\bibinfo{journal}{Nature Phys.}} \textbf{\bibinfo{volume}{4}},
  \bibinfo{pages}{463--466} (\bibinfo{year}{2008}).

\bibitem{lanyon2011universal}
\bibinfo{author}{Lanyon, B.~P.} \emph{et~al.}
\newblock \bibinfo{title}{Universal digital quantum simulation with trapped
  ions}.
\newblock \emph{\bibinfo{journal}{Science}} \textbf{\bibinfo{volume}{334}},
  \bibinfo{pages}{57--61} (\bibinfo{year}{2011}).

\bibitem{eckstein2010dielectric}
\bibinfo{author}{Eckstein, M.}, \bibinfo{author}{Oka, T.} \&
  \bibinfo{author}{Werner, P.}
\newblock \bibinfo{title}{Dielectric breakdown of mott insulators in dynamical
  mean-field theory}.
\newblock \emph{\bibinfo{journal}{Phys. Rev. Lett.}}
  \textbf{\bibinfo{volume}{105}}, \bibinfo{pages}{146404}
  (\bibinfo{year}{2010}).

\bibitem{eckstein2011damping}
\bibinfo{author}{Eckstein, M.} \& \bibinfo{author}{Werner, P.}
\newblock \bibinfo{title}{Damping of {B}loch oscillations in the {H}ubbard
  model}.
\newblock \emph{\bibinfo{journal}{Phys. Rev. Lett.}}
  \textbf{\bibinfo{volume}{107}}, \bibinfo{pages}{186406}
  (\bibinfo{year}{2011}).

\bibitem{PhysRevLett.103.056403}
\bibinfo{author}{Eckstein, M.}, \bibinfo{author}{Kollar, M.} \&
  \bibinfo{author}{Werner, P.}
\newblock \bibinfo{title}{Thermalization after an interaction quench in the
  {H}ubbard model}.
\newblock \emph{\bibinfo{journal}{Phys. Rev. Lett.}}
  \textbf{\bibinfo{volume}{103}}, \bibinfo{pages}{056403}
  (\bibinfo{year}{2009}).

\bibitem{PhysRevB.81.115131}
\bibinfo{author}{Eckstein, M.}, \bibinfo{author}{Kollar, M.} \&
  \bibinfo{author}{Werner, P.}
\newblock \bibinfo{title}{Interaction quench in the {H}ubbard model: Relaxation
  of the spectral function and the optical conductivity}.
\newblock \emph{\bibinfo{journal}{Phys. Rev. B}} \textbf{\bibinfo{volume}{81}},
  \bibinfo{pages}{115131} (\bibinfo{year}{2010}).

\bibitem{wall2011quantum}
\bibinfo{author}{Wall, S.} \emph{et~al.}
\newblock \bibinfo{title}{Quantum interference between charge excitation paths
  in a solid-state {M}ott insulator}.
\newblock \emph{\bibinfo{journal}{Nature Phys.}} \textbf{\bibinfo{volume}{7}},
  \bibinfo{pages}{114--118} (\bibinfo{year}{2011}).

\bibitem{fausti2011light}
\bibinfo{author}{Fausti, D.} \emph{et~al.}
\newblock \bibinfo{title}{Light-induced superconductivity in a stripe-ordered
  cuprate}.
\newblock \emph{\bibinfo{journal}{Science}} \textbf{\bibinfo{volume}{331}},
  \bibinfo{pages}{189--191} (\bibinfo{year}{2011}).

\bibitem{cardy1996scaling}
\bibinfo{author}{Cardy, J.}
\newblock \emph{\bibinfo{title}{Scaling and Renormalization in Statistical
  Physics}}, vol.~\bibinfo{volume}{5} (\bibinfo{publisher}{Cambridge University
  Press}, \bibinfo{year}{1996}).

\bibitem{RevModPhys.77.1027}
\bibinfo{author}{Maier, T.}, \bibinfo{author}{Jarrell, M.},
  \bibinfo{author}{Pruschke, T.} \& \bibinfo{author}{Hettler, M.~H.}
\newblock \bibinfo{title}{Quantum cluster theories}.
\newblock \emph{\bibinfo{journal}{Rev. Mod. Phys.}}
  \textbf{\bibinfo{volume}{77}}, \bibinfo{pages}{1027--1080}
  (\bibinfo{year}{2005}).

\bibitem{PhysRevB.90.075117}
\bibinfo{author}{Tsuji, N.}, \bibinfo{author}{Barmettler, P.},
  \bibinfo{author}{Aoki, H.} \& \bibinfo{author}{Werner, P.}
\newblock \bibinfo{title}{Nonequilibrium dynamical cluster theory}.
\newblock \emph{\bibinfo{journal}{Phys. Rev. B}} \textbf{\bibinfo{volume}{90}},
  \bibinfo{pages}{075117} (\bibinfo{year}{2014}).

\bibitem{gramsch2013hamiltonian}
\bibinfo{author}{Gramsch, C.}, \bibinfo{author}{Balzer, K.},
  \bibinfo{author}{Eckstein, M.} \& \bibinfo{author}{Kollar, M.}
\newblock \bibinfo{title}{Hamiltonian-based impurity solver for nonequilibrium
  dynamical mean-field theory}.
\newblock \emph{\bibinfo{journal}{Phys. Rev. B}} \textbf{\bibinfo{volume}{88}},
  \bibinfo{pages}{235106} (\bibinfo{year}{2013}).

\bibitem{PhysRevB.90.235131}
\bibinfo{author}{Wolf, F.~A.}, \bibinfo{author}{McCulloch, I.~P.} \&
  \bibinfo{author}{Schollw\"ock, U.}
\newblock \bibinfo{title}{Solving nonequilibrium dynamical mean-field theory
  using matrix product states}.
\newblock \emph{\bibinfo{journal}{Phys. Rev. B}} \textbf{\bibinfo{volume}{90}},
  \bibinfo{pages}{235131} (\bibinfo{year}{2014}).

\bibitem{cirac2012goals}
\bibinfo{author}{Cirac, J.~I.} \& \bibinfo{author}{Zoller, P.}
\newblock \bibinfo{title}{Goals and opportunities in quantum simulation}.
\newblock \emph{\bibinfo{journal}{Nature Phys.}} \textbf{\bibinfo{volume}{8}},
  \bibinfo{pages}{264--266} (\bibinfo{year}{2012}).

\bibitem{PhysRevB.91.045136}
\bibinfo{author}{Balzer, K.}, \bibinfo{author}{Li, Z.},
  \bibinfo{author}{Vendrell, O.} \& \bibinfo{author}{Eckstein, M.}
\newblock \bibinfo{title}{Multiconfiguration time-dependent {H}artree impurity
  solver for nonequilibrium dynamical mean-field theory}.
\newblock \emph{\bibinfo{journal}{Phys. Rev. B}} \textbf{\bibinfo{volume}{91}},
  \bibinfo{pages}{045136} (\bibinfo{year}{2015}).

\bibitem{dorner2013extracting}
\bibinfo{author}{Dorner, R.} \emph{et~al.}
\newblock \bibinfo{title}{Extracting quantum work statistics and fluctuation
  theorems by single-qubit interferometry}.
\newblock \emph{\bibinfo{journal}{Phys. Rev. Lett.}}
  \textbf{\bibinfo{volume}{110}}, \bibinfo{pages}{230601}
  (\bibinfo{year}{2013}).

\bibitem{casanova2012quantum}
\bibinfo{author}{Casanova, J.}, \bibinfo{author}{Mezzacapo, A.},
  \bibinfo{author}{Lamata, L.} \& \bibinfo{author}{Solano, E.}
\newblock \bibinfo{title}{Quantum simulation of interacting fermion lattice
  models in trapped ions}.
\newblock \emph{\bibinfo{journal}{Phys. Rev. Lett.}}
  \textbf{\bibinfo{volume}{108}}, \bibinfo{pages}{190502}
  (\bibinfo{year}{2012}).

\bibitem{muller2011simulating}
\bibinfo{author}{M{\"u}ller, M.}, \bibinfo{author}{Hammerer, K.},
  \bibinfo{author}{Zhou, Y.~L.}, \bibinfo{author}{Roos, C.~F.} \&
  \bibinfo{author}{Zoller, P.}
\newblock \bibinfo{title}{Simulating open quantum systems: from many-body
  interactions to stabilizer pumping}.
\newblock \emph{\bibinfo{journal}{New J. Phys.}} \textbf{\bibinfo{volume}{13}},
  \bibinfo{pages}{085007} (\bibinfo{year}{2011}).

\bibitem{molmer1999multiparticle}
\bibinfo{author}{M{\o}lmer, K.} \& \bibinfo{author}{S{\o}rensen, A.}
\newblock \bibinfo{title}{Multiparticle entanglement of hot trapped ions}.
\newblock \emph{\bibinfo{journal}{Phys. Rev. Lett.}}
  \textbf{\bibinfo{volume}{82}}, \bibinfo{pages}{1835} (\bibinfo{year}{1999}).

\bibitem{esslinger2010}
\bibinfo{author}{Esslinger, T.}
\newblock \bibinfo{title}{Fermi–{H}ubbard physics with atoms in an optical
  lattice}.
\newblock \emph{\bibinfo{journal}{Annu. Rev. Condens. Matter Phys.}}
  \textbf{\bibinfo{volume}{1}}, \bibinfo{pages}{129} (\bibinfo{year}{2010}).

\bibitem{langen2015ultracold}
\bibinfo{author}{Langen, T.}, \bibinfo{author}{Geiger, R.} \&
  \bibinfo{author}{Schmiedmayer, J.}
\newblock \bibinfo{title}{Ultracold atoms out of equilibrium}.
\newblock \emph{\bibinfo{journal}{Annu. Rev. Condens. Matter Phys.}}
  \textbf{\bibinfo{volume}{6}}, \bibinfo{pages}{201--217}
  (\bibinfo{year}{2015}).

\bibitem{PhysRevA.69.042314}
\bibinfo{author}{Fowler, A.~G.}, \bibinfo{author}{Hill, C.~D.} \&
  \bibinfo{author}{Hollenberg, L. C.~L.}
\newblock \bibinfo{title}{Quantum-error correction on linear-nearest-neighbor
  qubit arrays}.
\newblock \emph{\bibinfo{journal}{Phys. Rev. A}} \textbf{\bibinfo{volume}{69}},
  \bibinfo{pages}{042314} (\bibinfo{year}{2004}).

\bibitem{harty2014high}
\bibinfo{author}{Harty, T.~P.} \emph{et~al.}
\newblock \bibinfo{title}{High-fidelity preparation, gates, memory, and readout
  of a trapped-ion quantum bit}.
\newblock \emph{\bibinfo{journal}{Phys. Rev. Lett.}}
  \textbf{\bibinfo{volume}{113}}, \bibinfo{pages}{220501}
  (\bibinfo{year}{2014}).

\bibitem{superfermion}
\bibinfo{author}{Dzhioev, A.~A.} \& \bibinfo{author}{Kosov, D.~S.}
\newblock \bibinfo{title}{Super-fermion representation of quantum kinetic
  equations for the electron transport problem}.
\newblock \emph{\bibinfo{journal}{J. Chem. Phys.}}
  \textbf{\bibinfo{volume}{134}}, \bibinfo{pages}{044121}
  (\bibinfo{year}{2011}).

\bibitem{bauer}
\bibinfo{author}{Bauer, B.}, \bibinfo{author}{Wecker, D.},
  \bibinfo{author}{Millis, A.~J.}, \bibinfo{author}{Hastings, M.~B.} \&
  \bibinfo{author}{Troyer, M.}
\newblock \bibinfo{title}{Hybrid quantum-classical approach to correlated
  materials}.
\newblock \emph{\bibinfo{journal}{arXiv:1510.03859}}  (\bibinfo{year}{2015}).

\end{thebibliography}

\begin{thebibliography}{10}
\expandafter\ifx\csname url\endcsname\relax
  \def\url#1{\texttt{#1}}\fi
\expandafter\ifx\csname urlprefix\endcsname\relax\def\urlprefix{URL }\fi
\providecommand{\bibinfo}[2]{#2}
\providecommand{\eprint}[2][]{\url{#2}}

\bibitem{wall2011quantumSup}
\bibinfo{author}{Wall, S.} \emph{et~al.}
\newblock \bibinfo{title}{Quantum interference between charge excitation paths
  in a solid-state {M}ott insulator}.
\newblock \emph{\bibinfo{journal}{Nature Phys.}} \textbf{\bibinfo{volume}{7}},
  \bibinfo{pages}{114--118} (\bibinfo{year}{2011}).

\bibitem{georges1996dynamicalSup}
\bibinfo{author}{Georges, A.}, \bibinfo{author}{Kotliar, G.},
  \bibinfo{author}{Krauth, W.} \& \bibinfo{author}{Rozenberg, M.~J.}
\newblock \bibinfo{title}{Dynamical mean-field theory of strongly correlated
  fermion systems and the limit of infinite dimensions}.
\newblock \emph{\bibinfo{journal}{Rev. Mod. Phys.}}
  \textbf{\bibinfo{volume}{68}}, \bibinfo{pages}{13} (\bibinfo{year}{1996}).

\bibitem{RevModPhys.86.779Sup}
\bibinfo{author}{Aoki, H.} \emph{et~al.}
\newblock \bibinfo{title}{Nonequilibrium dynamical mean-field theory and its
  applications}.
\newblock \emph{\bibinfo{journal}{Rev. Mod. Phys.}}
  \textbf{\bibinfo{volume}{86}}, \bibinfo{pages}{779--837}
  (\bibinfo{year}{2014}).

\bibitem{gramsch2013hamiltonianSup}
\bibinfo{author}{Gramsch, C.}, \bibinfo{author}{Balzer, K.},
  \bibinfo{author}{Eckstein, M.} \& \bibinfo{author}{Kollar, M.}
\newblock \bibinfo{title}{Hamiltonian-based impurity solver for nonequilibrium
  dynamical mean-field theory}.
\newblock \emph{\bibinfo{journal}{Phys. Rev. B}} \textbf{\bibinfo{volume}{88}},
  \bibinfo{pages}{235106} (\bibinfo{year}{2013}).

\bibitem{PhysRevB.90.235131Sup}
\bibinfo{author}{Wolf, F.~A.}, \bibinfo{author}{McCulloch, I.~P.} \&
  \bibinfo{author}{Schollw\"ock, U.}
\newblock \bibinfo{title}{Solving nonequilibrium dynamical mean-field theory
  using matrix product states}.
\newblock \emph{\bibinfo{journal}{Phys. Rev. B}} \textbf{\bibinfo{volume}{90}},
  \bibinfo{pages}{235131} (\bibinfo{year}{2014}).

\bibitem{blatt2012quantumSup}
\bibinfo{author}{Blatt, R.} \& \bibinfo{author}{Roos, C.~F.}
\newblock \bibinfo{title}{Quantum simulations with trapped ions}.
\newblock \emph{\bibinfo{journal}{Nature Phys.}} \textbf{\bibinfo{volume}{8}},
  \bibinfo{pages}{277--284} (\bibinfo{year}{2012}).

\bibitem{casanova2012quantumSup}
\bibinfo{author}{Casanova, J.}, \bibinfo{author}{Mezzacapo, A.},
  \bibinfo{author}{Lamata, L.} \& \bibinfo{author}{Solano, E.}
\newblock \bibinfo{title}{Quantum simulation of interacting fermion lattice
  models in trapped ions}.
\newblock \emph{\bibinfo{journal}{Phys. Rev. Lett.}}
  \textbf{\bibinfo{volume}{108}}, \bibinfo{pages}{190502}
  (\bibinfo{year}{2012}).

\bibitem{muller2011simulatingSup}
\bibinfo{author}{M{\"u}ller, M.}, \bibinfo{author}{Hammerer, K.},
  \bibinfo{author}{Zhou, Y.~L.}, \bibinfo{author}{Roos, C.~F.} \&
  \bibinfo{author}{Zoller, P.}
\newblock \bibinfo{title}{Simulating open quantum systems: from many-body
  interactions to stabilizer pumping}.
\newblock \emph{\bibinfo{journal}{New J. Phys.}} \textbf{\bibinfo{volume}{13}},
  \bibinfo{pages}{085007} (\bibinfo{year}{2011}).

\bibitem{dorner2013extractingSup}
\bibinfo{author}{Dorner, R.} \emph{et~al.}
\newblock \bibinfo{title}{Extracting quantum work statistics and fluctuation
  theorems by single-qubit interferometry}.
\newblock \emph{\bibinfo{journal}{Phys. Rev. Lett.}}
  \textbf{\bibinfo{volume}{110}}, \bibinfo{pages}{230601}
  (\bibinfo{year}{2013}).

\bibitem{superfermionSup}
\bibinfo{author}{Dzhioev, A.~A.} \& \bibinfo{author}{Kosov, D.~S.}
\newblock \bibinfo{title}{Super-fermion representation of quantum kinetic
  equations for the electron transport problem}.
\newblock \emph{\bibinfo{journal}{J. Chem. Phys.}}
  \textbf{\bibinfo{volume}{134}}, \bibinfo{pages}{044121}
  (\bibinfo{year}{2011}).

\end{thebibliography}
\end{document}